\documentclass[epj]{svjour}
\usepackage{graphics}
\usepackage{epstopdf}

\begin{document}

\newcommand{\vk}{{\vec k}}
\newcommand{\vK}{{\vec K}}
\newcommand{\vb}{{\vec b}}
\newcommand{{\vp}}{{\vec p}}
\newcommand{{\vq}}{{\vec q}}
\newcommand{\vQ}{{\vec Q}}
\newcommand{\vx}{{\vec x}}
\newcommand{\beq}{\begin{equation}}
\newcommand{\eeq}{\end{equation}}
\newcommand{\half}{{\textstyle \frac{1}{2}}}
\newcommand{\gton}{\stackrel{>}{\sim}}
\newcommand{\lton}{\mathrel{\lower.9ex \hbox{$\stackrel{\displaystyle<}{\sim}$}}}
\newcommand{\ee}{\end{equation}}
\newcommand{\ben}{\begin{enumerate}}
\newcommand{\een}{\end{enumerate}}
\newcommand{\bit}{\begin{itemize}}
\newcommand{\eit}{\end{itemize}}
\newcommand{\bc}{\begin{center}}
\newcommand{\ec}{\end{center}}
\newcommand{\bea}{\begin{eqnarray}}
\newcommand{\eea}{\end{eqnarray}}

\newcommand{\beqar}{\begin{eqnarray}}
\newcommand{\eeqar}[1]{\label{#1} \end{eqnarray}}
\newcommand{\pleft}{\stackrel{\leftarrow}{\partial}}
\newcommand{\pright}{\stackrel{\rightarrow}{\partial}}

\newcommand{\eq}[1]{Eq.~(\ref{#1})}
\newcommand{\fig}[1]{Fig.~\ref{#1}}
\newcommand{\eff}{ef\!f}
\newcommand{\alphas}{\alpha_s}

\renewcommand{\topfraction}{0.85}
\renewcommand{\textfraction}{0.1}
\renewcommand{\floatpagefraction}{0.75}

\title{NLO Productions of $\omega$ and $K^0_{\rm S}$ with a Global Extraction of the Jet Transport Parameter in Heavy Ion collisions}
\author{Guo-Yang Ma\inst{1} \and Wei Dai\inst{2}\thanks{ weidai@cug.edu.cn} \and Ben-Wei Zhang\inst{1} \thanks{ bwzhang@mail.ccnu.edu.cn} \and Enke Wang\inst{1}
%
%
}                     
%
%
\institute{Key Laboratory of Quark \& Lepton Physics (MOE) and Institute of Particle Physics,
 Central China Normal University, Wuhan 430079, China \and School of Mathematics and Physics, China University of Geosciences, Wuhan 430079, China}

\date{Received: date / Revised version: date}
%
\abstract{
In this work, we pave the way to calculate the productions of $\omega$ and $K^0_{\rm S}$ mesons with large $p_T$ in p+p and A+A collisions both at RHIC and LHC. The fragmentation functions (FFs) of $\omega$ meson in vacuum at next-to-leading order (NLO) are obtained by evolving NLO DGLAP evolution equations with rescaled $\omega$ FFs at initial scale $Q_0^2=1.5$~GeV$^2$ from a broken SU(3) model, and the FFs of $K^0_{\rm S}$ in vacuum are taken from AKK08 parametrization directly. Within the framework of the NLO pQCD improved parton model,
we make good descriptions of the experimental data on $\omega$ and $K^0_{\rm S}$ in p+p both at RHIC and LHC. With the higher-twist approach to take into account jet quenching effect by medium modified FFs, nuclear modification factors for $\omega$ meson and $K^0_{\rm S}$ meson both at RHIC and LHC are presented with different sets of jet transport coefficient $\hat{q}_0$. Then we make a global extraction of $\hat{q}_0$ both at RHIC and LHC by confronting our model calculations with all available data on 6 identified mesons:
$\pi^0$, $\eta$, $\rho^0$, $\phi$, $\omega$, and $K^0_{\rm S}$. The minimum value of total $\chi^2/d.o.f$ for productions of these mesons gives the best value of $\hat{q}_0=0.5\rm~GeV^2/fm$ for Au+Au collisions with $\sqrt{s_{\rm NN}}=200$~GeV at RHIC, and $\hat{q}_0=1.2\rm~GeV^2/fm$ for Pb+Pb collisions with $\sqrt{s_{\rm NN}}=2.76$~TeV at LHC respectively, with the QGP spacetime evolution given by an event-by-event viscous hydrodynamics model IEBE-VISHNU. With these global extracted values of $\hat{q}_0$, nuclear modification factors of $\pi^0$, $\eta$, $\rho^0$, $\phi$, $\omega$, and $K^0_{\rm S}$ in A+A collisions are presented, and predictions of yield ratios such as $\omega/\pi^0$ and $K^0_{\rm S}/\pi^0$ at high-$p_T$ regime in heavy-ion collisions both at RHIC and LHC are provided.
} 
\titlerunning{ $\omega$ and $K^0_{\rm S}$ Productions with a Global Extraction of the Jet Transport Parameter}
\maketitle
\section{Introduction}
\label{intro}

Jet quenching effect describes energy dissipation of an energetic parton when it traverses through the hot and dense QCD medium, which is produced shortly after high energy nuclear collisions~\cite{Wang:1991xy}. The single hadron production suppression at high-$p_T$ regime when compared to scaled p+p data is one of the primary perturbative probes to study the properties of this de-coupled quark and gluon QCD matter~\cite{Gyulassy:2003mc}. $\pi^0$ as the most well measured final state hadron, its nuclear modification factor $R_{\rm AA}$ as a function of transverse momentum $p_T$ is interpreted as the consequence of jet quenching effect and it did help us to constrain the strength of jet-medium interaction and also the properties of QCD medium~\cite{Adler:2006hu,Adare:2008qa,Agakishiev:2011dc,Adare:2013esx,Abelev:2014laa,Acharya:2017hyu}. In 2013, Jet collaboration summarized different jet quenching theoretical frameworks and compared the jet transport parameter extracted by different energy loss models using the same hydro description of QCD medium~\cite{Burke:2013yra}. Meanwhile, the $R_{\rm AA}$ for different final state identified hadrons have been measured, and their production suppressions as well as patterns of their yield ratios have been observed~\cite{Adare:2008qa,Agakishiev:2011dc,Adare:2010dc,Adare:2010pt,Adam:2015kca,Acharya:2018yhg,Adare:2011ht,Adam:2017zbf}. It is of interest and challenge to describe the cross sections of leading hadron of different types and their yield ratios with each other in heavy-ion collisions (HIC) both at RHIC and LHC with a unified model of jet quenching, which should shed light on flavor dependence of parton energy loss and the intrinsic properties of identified hadron productions in p+p and A+A reactions
\cite{Abelev:2006jr,Xu:2010fs,Liu:2006sf,Brodsky:2006ha,Chen:2008vha,Chen:2010te,Chen:2011vt,Liu:2015vna}.

We have achieved more understanding of the suppression patterns of different mesons by conducting calculations and analysis of the $R_{\rm AA}$ and particle ratios for other identified hadron productions~\cite{Dai:2015dxa,Dai:2017tuy,Dai:2017piq}. We find the understanding of the suppression pattern of the leading hadron requires to take into account all three factors~\cite{Dai:2015dxa}: the initial hard jet spectrum, the energy loss mechanism and the parton fragmentation functions in vacuum. The flavour dependence of the energy loss will result in the decreasing the fraction of gluon fragmenting contribution in the nuclear-nuclear collision and increasing the fraction of quark fragmenting contribution. The productions of final state mesons, like $\pi^0$,$\rho^0$,$\eta$~\cite{Dai:2015dxa,Dai:2017tuy,Dai:2017piq}, are dominated by the quark fragmentation contribution at high-$p_T$ regime in p+p collision. The energy loss effect will only enhance the domination of quark fragmenting contribution in A+A collision. Therefore both $\rho^0/\pi^0$ and $\eta/\pi^0$ in p+p and A+A collision are coincided at large-$p_T$ since they are only determined by the ratios of the FFs in vacuum. But the production of $\phi$ meson is dominated by gluon fragmenting contribution in p+p collision, therefore we can observe a separation of $\phi/\pi^0$ in p+p and A+A. It is of great interest to gain more understanding of the suppression pattern of different final state hadrons, and further provide systematical pQCD predictions for the current experimental measurement of identified hadrons. In this letter, we mainly investigate one of low-mass vector mesons ($\omega$) and one of pseudoscalar mesons ($K_S^0$), together with $\pi^0$, $\eta$, $\rho^0$ and $\phi$ to achieve these goals.

In this article, we investigate the production of other two mesons $\omega$ and $K^0_{\rm S}$ with large $p_T$ in A+A collisions, which has never been computed so far, to the best of our knowledge. The $\omega$ meson is constituted of similar valence quark of $\pi^0$ with larger mass~$782.65$~MeV and spin $1$. Kaons are a group of lightest mesons, carrying strangeness components, $K^0_{\rm S}(\rm\frac{d\bar{s}+s\bar{d}}{2})$ is one type of the Kaons, which is consisted by $\rm s-$quark, $\rm d-$quark and their corresponding anti-quarks. The productions of these two mesons have been measured in p+p and A+A collisions both at RHIC and LHC, but lack the theoretical description. Within the NLO pQCD improved parton model, we calculate $\omega$ and $K^0_{\rm S}$ yields at high $p_T$ regime in heavy-ion collisions, by employing medium modification fragmentation functions (FFs) due to gluon radiation in the hot/dense QCD meidum in the higher-twist approach of jet quenching~\cite{Chen:2010te,Chen:2011vt,Zhang:2003yn,Zhang:2003wk,Schafer:2007xh}, the same approach as in our calculations on productions of $\eta$, $\rho^0$ and $\phi$ mesons in HIC~\cite{Dai:2015dxa,Dai:2017tuy,Dai:2017piq}.

It is noted in the previous studies, for the consistency, we implemented Hirano hydro description~\cite{Hirano:2001eu,Hirano:2002ds} to describe the space-time evolution of the QGP fireball. In this work, we utilize a state-of-art, event-by-event (2+1)-D viscous hydrodynamics model (IEBE-VISHNU)~\cite{Shen:2014vra} to give the space-time evolution information of the hot and dense medium. Due to the changes on the medium description, the strength of jet-medium interaction characterized by the jet transport parameter $\hat{q}_0$ should be re-extracted. Taking the advantage of systematical study on the nuclear modification factors with respect to $p_T$ and a large amount of  experimental data of $\pi^0$, $\eta$, $\rho^0$ and $\phi$ both at RHIC and LHC, with two more mesons $\omega$ and $K^0_{\rm S}$ calculated in this article, we can make a global extraction of the jet transport parameter $\hat{q}_0$ with all the available experimental data of these 6 identified mesons $R_{\rm AA}$.

The article is organized as follows. We first present the theoretical framework of computing single hadron cross sections in p+p collision in Sec. 2, and give the p+p baseline in investigating the in-medium modification of these productions. In Sec. 3, we discuss the inclusive hadron production in A+A collisions with the medium-modified FFs, basing on the higher-twist approach of parton energy loss, and present the nuclear modification factors $R_{\rm AA}$ for $\omega$ and $K^0_{\rm S}$ both at RHIC and LHC. Sec. 4 shows a global extraction of jet transport coefficient $\hat{q}_0$ both at RHIC and LHC by confronting our model calculations with all available data on 6 identified mesons:  $\pi^0$, $\eta$, $\rho^0$,  $\phi$,  $\omega$, and $K^0_{\rm S}$.  In Sec.5, we make predictions on the identified hadron yield ratios $\omega/\pi^0$ and $K^0_{\rm S}/\pi^0$ in p+p and A+A collisions, and compare them with experimental data if applicable. We give a brief summary in Sec. 6.

\section{Large $p_{\rm T}$ yield of $\omega$ and $K^0_{\rm S}$ meson in p+p}
\label{p+p}

We start with the productions of $\omega$ and $K^0_{\rm S}$ mesons at NLO in p+p collisions. In the framework of the pQCD improved parton model, the  inclusive hadron production in p+p collision can be given by the convolution of  three parts,  parton distribution functions (PDFs) in a proton, hard partonic scattering cross section denoted as $d\sigma/d\hat{t}$ (up to the order of $\alpha_s^3$),  and the parton fragmentation functions (FFs) to the final state hadron $D_{q(g)\to h} (z_h, Q^2)$. One may convolute PDFs and partonic cross section $d\sigma/d\hat{t}$ into the initial hard (parton-) jet spectrum $F_{q,g}(p_T)$, and we have:
\begin{eqnarray}
\frac{1}{p_{T}}\frac{d\sigma_{\omega, K^0_{\rm S}}}{dp_{T}}=\int F_{q}(\frac{p_{T}}{z_{h}})\cdot D_{q\to \omega, K^0_{\rm S}}(z_{h}, p_{T})\frac{dz_{h}}{z_{h}^2} \nonumber  \\
+ \int F_{g}(\frac{p_{T}}{z_{h}})\cdot D_{g\to \omega, K^0_{\rm S}}(z_{h}, p_{T})\frac{dz_{h}}{z_{h}^2}  \,\,\, .
\label{eq:ptspec}
\end{eqnarray}
in which the quark and gluon fragmenting contributions are written separately to facilitate future
discussions. A next-to-leading order (NLO) Monte Carlo code has been employed to calculate leading hadron productions in p+p collision~\cite{Kidonakis:2000gi},  and the CT14 parametrization of PDFs for free proton ~\cite{Hou:2016sho} has been implemented. In the framework, as long as the parametrization of the specific final state hadron FFs in vacuum is available, one can predict its production yield at large $p_{\rm T}$
in elementary p+p collision in principle.
\begin{figure}[!b]
\begin{center}
\hspace*{-0.1in}
\vspace*{-0.1in}
\resizebox{0.53\textwidth}{!}{%
\includegraphics{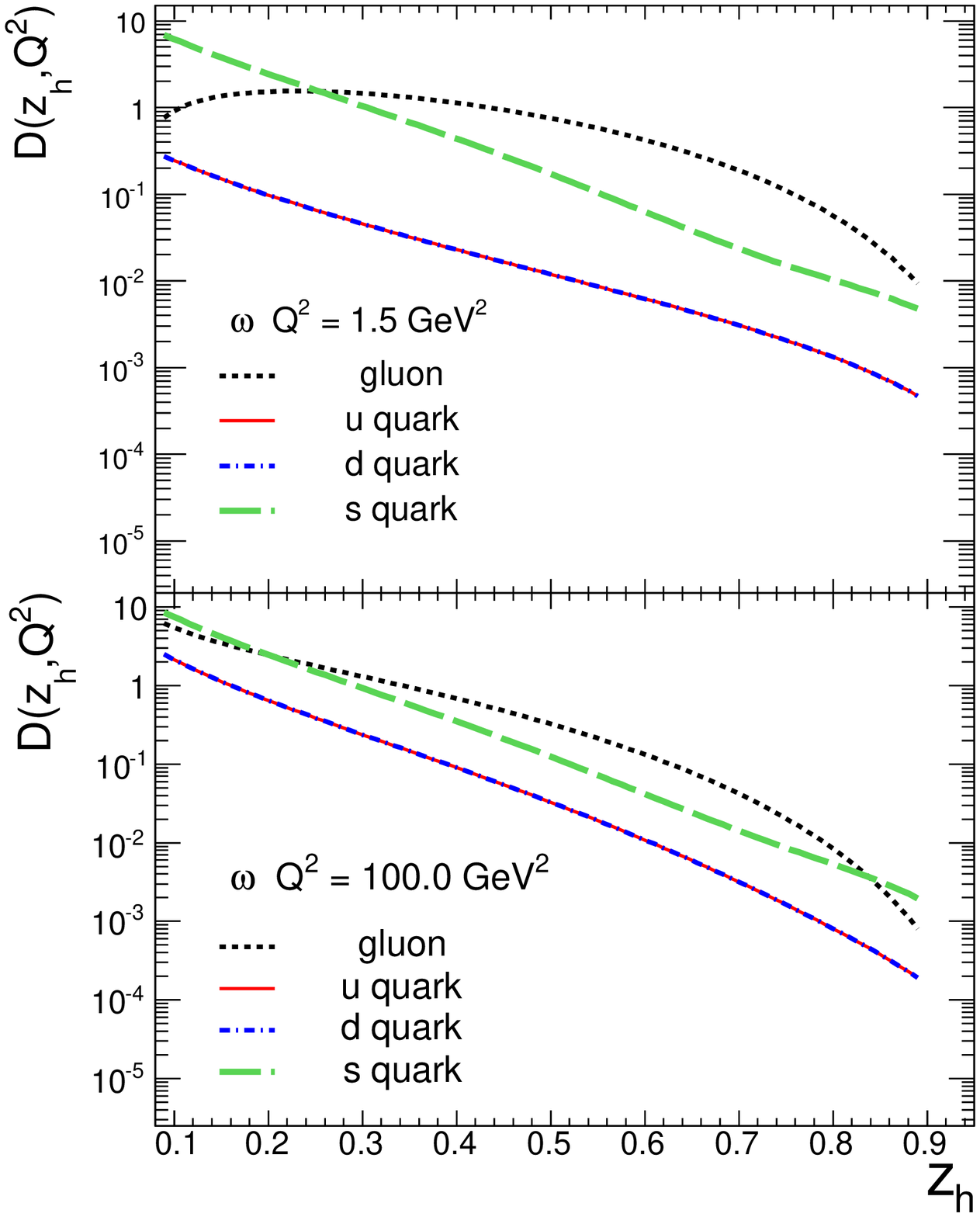}
\includegraphics{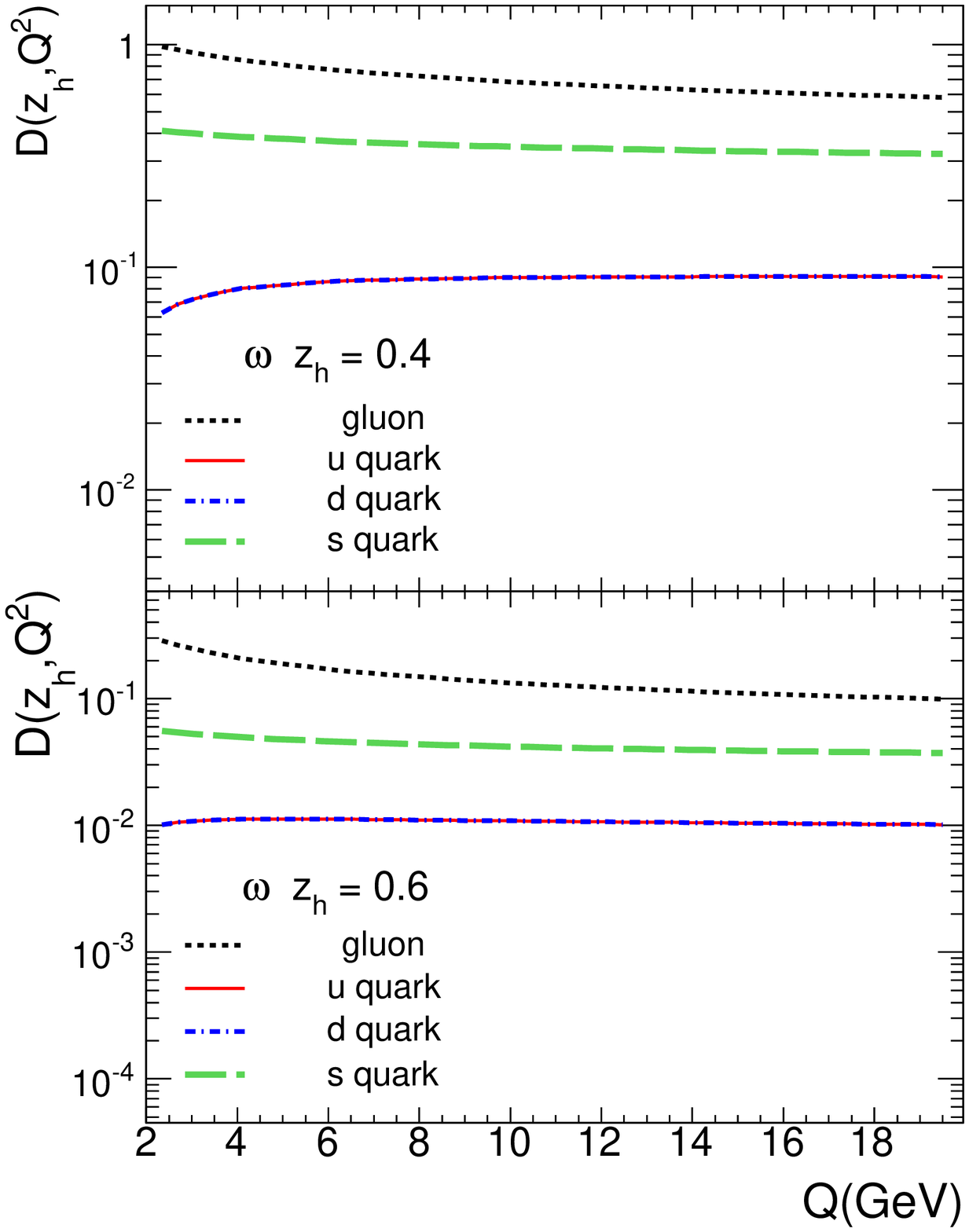}
}

\hspace*{-0.1in}
\vspace*{0.0in}
\caption{Left: The NLO DGLAP evolved Fragmentation Functions of $\omega$ meson as functions of $z_h$ at fixed $Q^2=1.5~\rm GeV^2$ and $Q^2=100~\rm GeV^2$. Right: The NLO DGLAP evolved Fragmentation Functions of $\omega$ meson as functions of $Q$ at fixed $z_h=0.4$ and $z_h=0.6$.}
\label{fig:omffs}
\end{center}
\end{figure}
\begin{figure}[!b]
\begin{center}
\hspace*{-0.1in}
\vspace*{-0.1in}
\resizebox{0.53\textwidth}{!}{%
\includegraphics{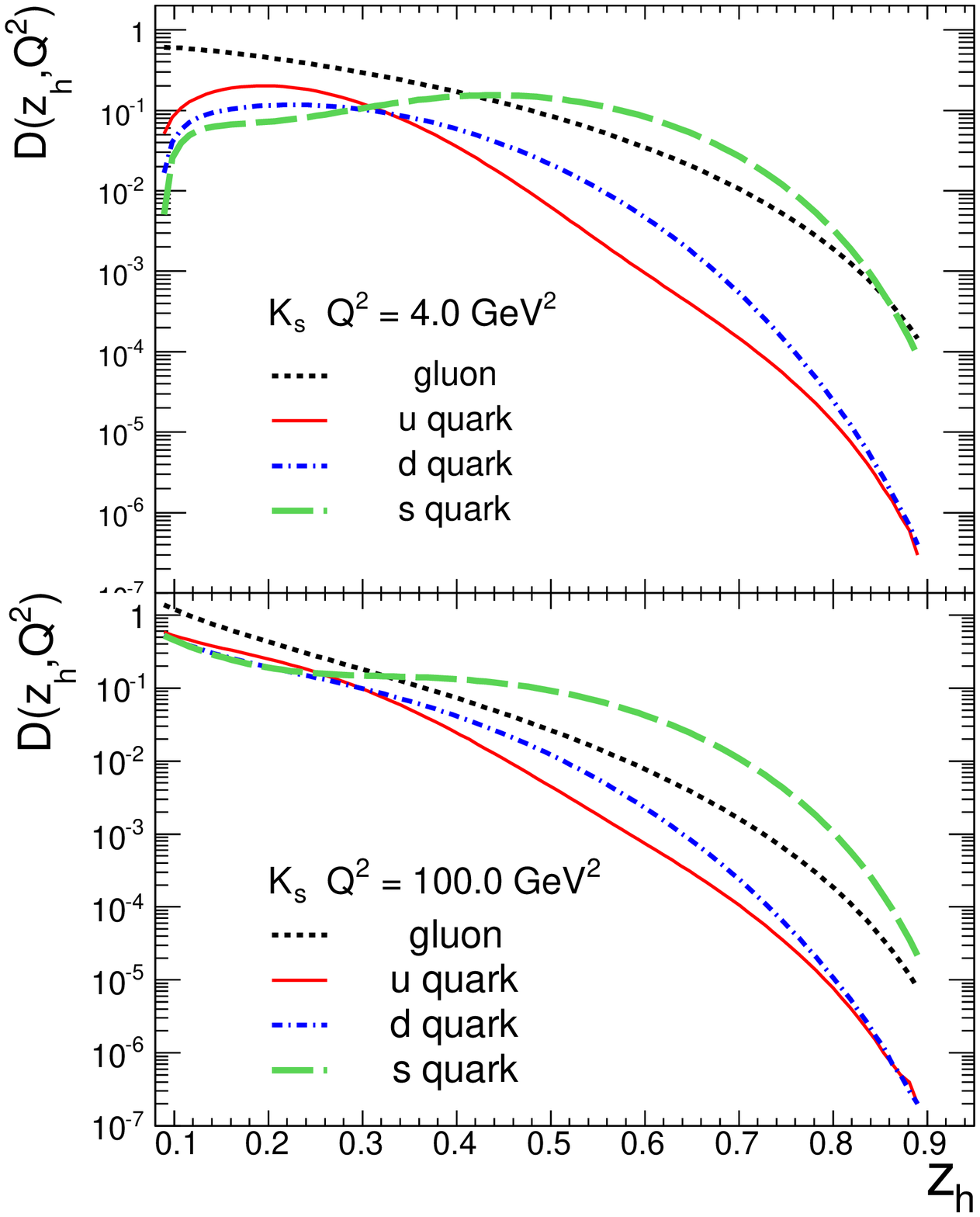}
\includegraphics{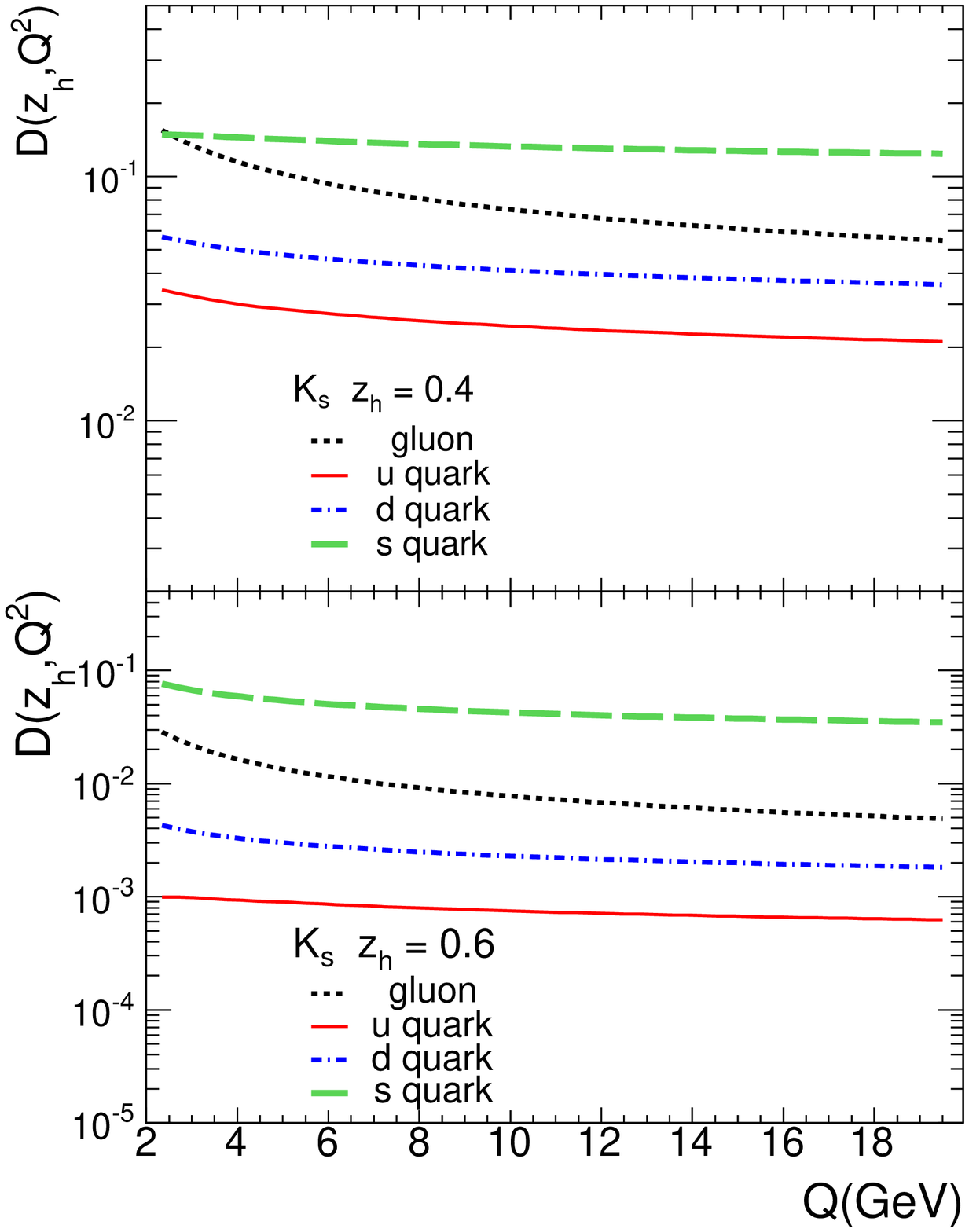}
}

\hspace*{-0.1in}
\vspace*{0.0in}
\caption{Left: The NLO DGLAP evolved Fragmentation Functions of $K^0_{\rm S}$ meson as functions of $z_h$ at fixed $Q^2=4.0~\rm GeV^2$ and $Q^2=100~\rm GeV^2$. Right: The NLO DGLAP evolved Fragmentation Functions of $K^0_{s}$ meson as functions of $Q$ at fixed $z_h=0.4 $ and $z_h=0.6$.}
\label{fig:ksffs}
\end{center}
\end{figure}

In order to make the NLO calculation of $\omega$ and $K^0_{\rm S}$ mesons in p+p, the NLO parton FFs of  these two mesons are needed. The parton FFs of $K^0_{\rm S}$ meson at NLO can be found in AKK08 parametrization~\cite{Albino:2008fy}, while there is not such kind of global parametrization for $\omega$ FFs in vacuum. So we have to rely on theoretical models and utilize $\omega$ parton FFs  at a starting scale $\rm Q_0^2=1.5~GeV^2$ provided by a broken SU(3) model~\cite{Indumathi:2011vn,Saveetha:2013jda}.  And we note that our preceding investigations on leading $\rho^0$ and $\phi$ production
~\cite{Dai:2017tuy,Dai:2017piq} have benefited from this broken SU(3) model. In this model, the independent parton FFs of different flavor are reduced into 3 independent functions named as valence $V(x, Q_0^2)$, sea $\gamma(x,Q_0^2)$ and gluon $D_g(x,Q_0^2)$ by considering the SU(3) flavor symmetry with a symmetry breaking parameter also the isospin and charge conjugation invariance of the vector mesons. We can write these three functions into a standard polynomial at the starting low energy scale of $\rm Q_0^2= 1.5~GeV^2$ as:
\begin{eqnarray}
H_i(x)=a_i x^{b_i}(1-x)^{c_i}(1+d_i x + e_i x^2)
\label{eq:form}
\end{eqnarray}

Where the full set of parameters($a,b,c,d,e$) defined in $V$, $\gamma$, $D_g$ and a few additional parameters defined for each vector meson such as strangeness suppression factor $\lambda$, vector mixing angle $\theta$, sea suppression factor $f_{sea}^\omega$, $f_1^u(\omega)$ and $f_g^\omega$ have been determined in the broken SU(3) model in Ref.~\cite{Indumathi:2011vn,Saveetha:2013jda}, and we multiple the parameter $a_i$ by three to make the best fit to the yield of $\omega$ meson in p+p collisions. All these parameters for fitting initial FFs of $\omega$ meson has been listed in Table.~\ref{tab:inputs}.
To obtain a NLO parton FFs for $\omega$ at any energy scale $Q$, we employ a numerical NLO DGLAP evolution program provided in Ref.~\cite{Hirai:2011si} with the initial parton FFs starting scale $\rm Q_0^2= 1.5~GeV^2$ as input.

We demonstrate in Fig.~\ref{fig:omffs} the NLO DGLAP evolved FFs of $\omega$ meson as functions of $z_h$ at fixed $\rm Q^2=1.5~GeV^2$ and $\rm Q^2=100~GeV^2$ on the left and as functions of $Q$ at fixed $z_h=0.4$ and $z_h=0.6$ on the right. We find in the typical fraction region ($z_h=0.4 \to 0.7$) $D_g^\omega > D_s^\omega \gg D_{u(d)}^\omega$, unlike the case $D_s^\phi > D_g^\phi \gg D_{u(d)}^\phi$ in the $\phi$ FFs~\cite{Dai:2017piq}. Due to the fact that in parton FFs, gluon exceeds quark contribution, an overwhelming advantage of gluon fragmenting contribution of the final state $\omega$ production in p+p collision is expected in the competition with quark fragmenting contribution.  In Fig.~\ref{fig:ksffs}, we plot the NLO FFs of $K^0_{\rm S}$ from AKK08 parametrization in the same manner as $\omega$ meson in Fig.~\ref{fig:omffs}. We find the strange quark FF $D_s^{K^0_{\rm S}}>D_g^{K^0_{\rm S}}$ in the typical $z_h=0.4 \to 0.7$ region, therefore a competition of the gluon and quark fragmenting contributions is expected in the $K^0_{\rm S}$ production in p+p collision based on the pattern discovered in the previous study on $\pi^0$, $\eta$, $\rho^0$ and $\phi$ meson.

\begin{figure}[!b]
\begin{center}

\resizebox{0.48\textwidth}{!}{%
\includegraphics{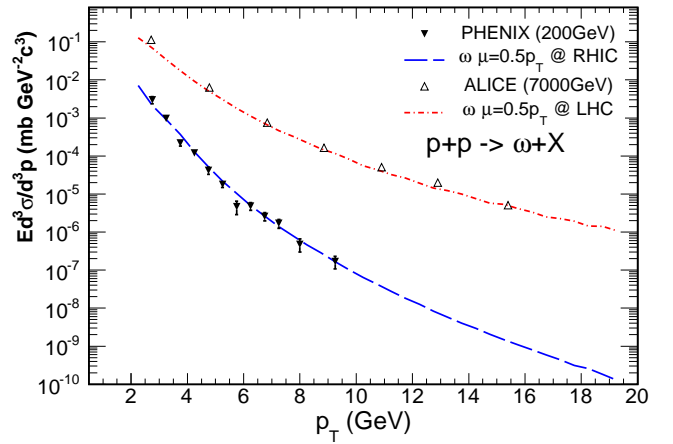}
}
\hspace*{-0.1in}
\caption{The theoretical results of the $\omega$ meson production in p+p collisions at $\sqrt s=200$~GeV confronted with the PHENIX data~\cite{Adare:2011ht} and confronted with the ALICE data~\cite{Peresunko:2012tt} at $\sqrt s=7$~TeV.
}
\label{fig:ompp}
\end{center}
\end{figure}

\begin{figure}[!t]
\begin{center}

\resizebox{0.48\textwidth}{!}{%
\includegraphics{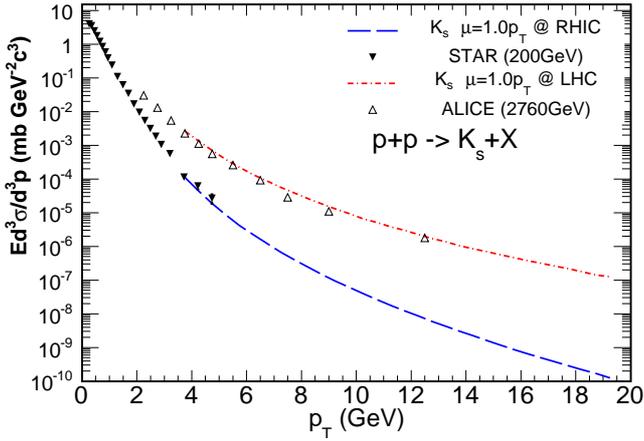}
}

\hspace*{-0.1in}
\caption{The theoretical results of the $K^0_{\rm s}$ meson production in p+p collisions at $\sqrt s=200$~GeV RHIC compared with STAR data~\cite{Abelev:2006cs} and at $\sqrt s=2.76$~TeV LHC compared with ALICE data~\cite{Adam:2017zbf}.
}
\label{fig:kspp}
\end{center}
\end{figure}

With the NLO FFs in vacuum of both $\omega$ and $K^0_{\rm S}$ meson, we are able to confront the numerical results of the $\omega$ and $K^0_{\rm S}$ meson productions in p+p collisions up to the NLO with the available experimental data both at RHIC and LHC shown in Fig.~\ref{fig:ompp} and Fig.~\ref{fig:kspp}. Note that in our simulations, the hard scales such as factorization scale, renormalization scale and fragmentation scale are chosen to be the same and proportional to the $p_T$ of the final state hadron.
We see the numerical results can match well with the experimental data on $\omega$ meson spectra both at RHIC $\sqrt{s_{\rm NN}}=200$~GeV and LHC $\sqrt{s_{\rm NN}}=7$~TeV and the hard scales $\mu=0.5$~$p_T$, as shown in Fig.~\ref{fig:ompp}. We also demonstrate in Fig.~\ref{fig:kspp} the good agreements of the theoretical calculations of leading $K^0_{\rm S}$ meson, with the STAR data and the ALICE data, where the hard scales are fixed to be $\mu=1.0$~$p_T$ to give the best fit.
\section{Large $p_{\rm T}$ yield of $\omega$ and $K^0_{\rm S}$ Meson in A+A}
\label{A+A}

To calculate the $\omega$ and $K^0_{\rm S}$ productions in A+A collisions, jet quenching effect should be included. In this article we utilize the higher-twist approach of parton energy loss, which relates  parton energy loss due to its multiple scattering in QCD medium and medium-induced gluon radiation to twist-four processes, and shows that these processes give rise to additional terms in QCD evolution equations and lead to the effectively medium-modified fragmentation functions; thus the partonic energy loss effect can be taken into account by replacing the FFs in vacuum with the effectively medium modified FFs~\cite{Zhang:2003yn,Zhang:2003wk,Schafer:2007xh,Guo:2000nz,Wang:2001ifa}. By assuming a thermal ensemble of quasi-particle states in the hot/dense medium and also neglecting the multiple particle correlations inside the medium, the effective medium modified FFs are given as~\cite{Chen:2010te,Chen:2011vt,Dai:2015dxa,Dai:2017tuy,Dai:2017piq}:
\begin{eqnarray}
\tilde{D}_{q}^{h}(z_h,Q^2) &=&
D_{q}^{h}(z_h,Q^2)+\frac{\alpha_s(Q^2)}{2\pi}
\int_0^{Q^2}\frac{d\ell_T^2}{\ell_T^2} \nonumber\\
&&\hspace{-0.7in}\times \int_{z_h}^{1}\frac{dz}{z} \left[ \Delta\gamma_{q\rightarrow qg}(z,x,x_L,\ell_T^2)D_{q}^h(\frac{z_h}{z},Q^2)\right.
\nonumber\\
&&\hspace{-0.2 in}+ \left. \Delta\gamma_{q\rightarrow
gq}(z,x,x_L,\ell_T^2)D_{g}^h(\frac{z_h}{z},Q^2) \right] .
\label{eq:mo-fragment}
\end{eqnarray}

The contribution of the medium-induced gluon radiation is attributed to the medium modified splitting functions represented by $\Delta\gamma_{q\rightarrow qg}$ and $\Delta\gamma_{q\rightarrow gq}$. In this formula, we assume the energy loss of the jet propagating through the medium is totally carried away by the radiative gluons, then the convolution of these energy loss kernels $\Delta\gamma_{q\rightarrow qg}$ ($\Delta\gamma_{q\rightarrow gq}$) with the (DGLAP) evolved vacuum FFs at any scale $D_{q,g}^h(\frac{z_h}{z},Q^2)$ implies the assumption that the fast parton first loses its energy in the medium and then fragments into final state hadrons in vacuum.

The medium modified splitting functions depend on the properties of the local medium which can not be determined directly by the theoretical calculation itself. The jet transport parameter $\hat{q}$ which defined as the average squared transverse momentum broadening per unit length is therefore introduced to profile the dependency of the local medium properties in these energy loss kernels~\cite{Burke:2013yra}. The jet transport parameter $\hat{q}$ can be phenomenological assumed to be proportional to the local parton density in the hot/dense medium to adopt further medium description:
\begin{equation}
\label{q-hat-qgph}
\hat{q} (\tau,r)= \hat{q}_0\frac{\rho_{\rm med}(\tau,r)}{\rho_{\rm med}(\tau_{0},0)}
  \cdot \frac{p^\mu u_\mu}{p_0}\,,
\end{equation}

Where $\rho_{\rm med}$ is the parton (quarks and gluon) density in the medium at a given temperature, $q_{0}$ is the initial jet transport parameter at the center of bulk medium at initial time $\tau_{0}$, $p^\mu$ is the four-momentum of jet and $u^\mu$ is the four flow velocity of the probed local medium in the collision frame. In this article we employ the state-of-art, event-by-event viscous hydrodynamics description IEBE-VISHNU~\cite{Shen:2014vra} to give the space-time evolution information of the hot/dense medium such as temperature, energy density, four velocity of the local medium at any evolution time and position,  as well as the formation time of QGP $\tau_0=0.6$~fm. Therefore the only undetermined parameter is the calculation is $\hat{q}_0$ representing the strength of jet-medium interaction, which can be fixed by fitting the data of leading hadrons (usually $\pi$ meson) suppression in A+A
collisions~\cite{Burke:2013yra,Dai:2015dxa,Dai:2017tuy,Dai:2017piq}.

When averaging the medium modified FFs over the initial production position and jet propagation direction, we are able to directly replace the vacuum FFs in the p+p formalism with these averaged medium modified FFs $\langle \tilde{D}_{c}^{h}(z_{h},Q^2,E,b)\rangle$, therefore the formalism of the calculated cross section of the single hadron productions in HIC would take the form as:
\begin{eqnarray}
\frac{1}{\langle N_{\rm coll}^{\rm AB}(b)\rangle}\frac{d\sigma_{AB}^h}{dyd^2p_T} &=&\sum_{abcd}\int
dx_adx_b f_{a/A}(x_a,\mu^2)f_{b/B}(x_b,\mu^2) \nonumber \\
&&\hspace{-0.5in}\times \frac{d\sigma}{d\hat{t}}(ab\rightarrow
cd)\frac{\langle \tilde{D}_{c}^{h}(z_{h},Q^2,E,b)\rangle}{\pi z_{c}}+\mathcal {O}(\alpha_s^3). \nonumber \\
\label{eq:AA}
\end{eqnarray}

In which  $\langle N_{\rm coll}^{\rm AB}(b)\rangle=\int d^{2}r t_{A}(r)t_{B}(|\vec b-\vec r|)$ is the number of binary nucleon-nucleon collisions at certain impact parameter $b$ in A+B collisions and it can be calculated by using Glauber Model~\cite{dEnterria:2003xac}. $f_{a/A}(x_{a},\mu^{2})$ represents the effective PDFs inside a nucleus. In our calculations, we employed EPPS16 NLO nuclear PDFs~\cite{Eskola:2016oht} to include initial-state cold nuclear matter effects on single hadron productions~\cite{Chen:2015qmd}.
\begin{figure}[!t]
\begin{center}
\hspace*{-0.1in}
\vspace*{-0.1in}
\resizebox{0.515\textwidth}{!}{%
\includegraphics{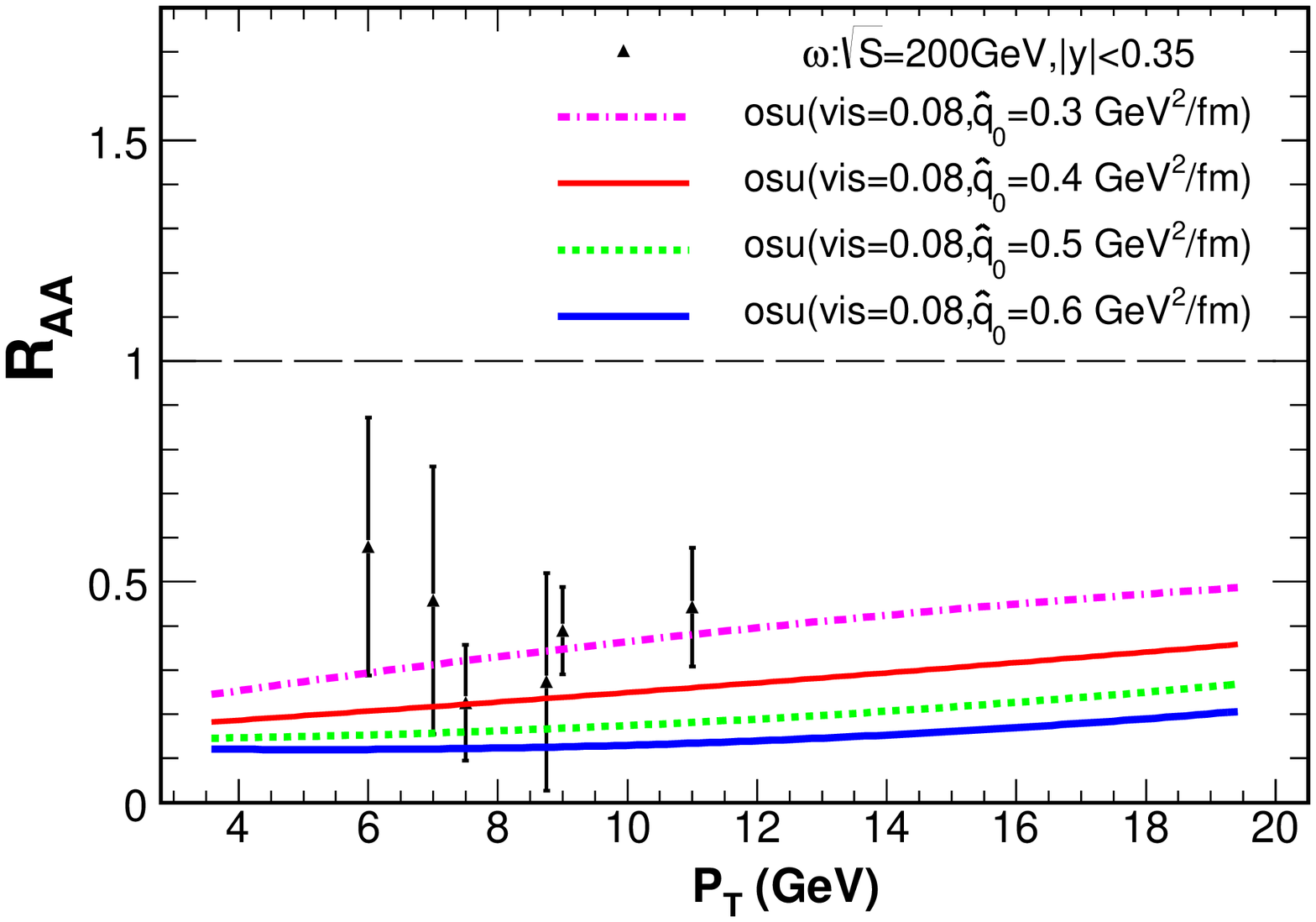}
\includegraphics{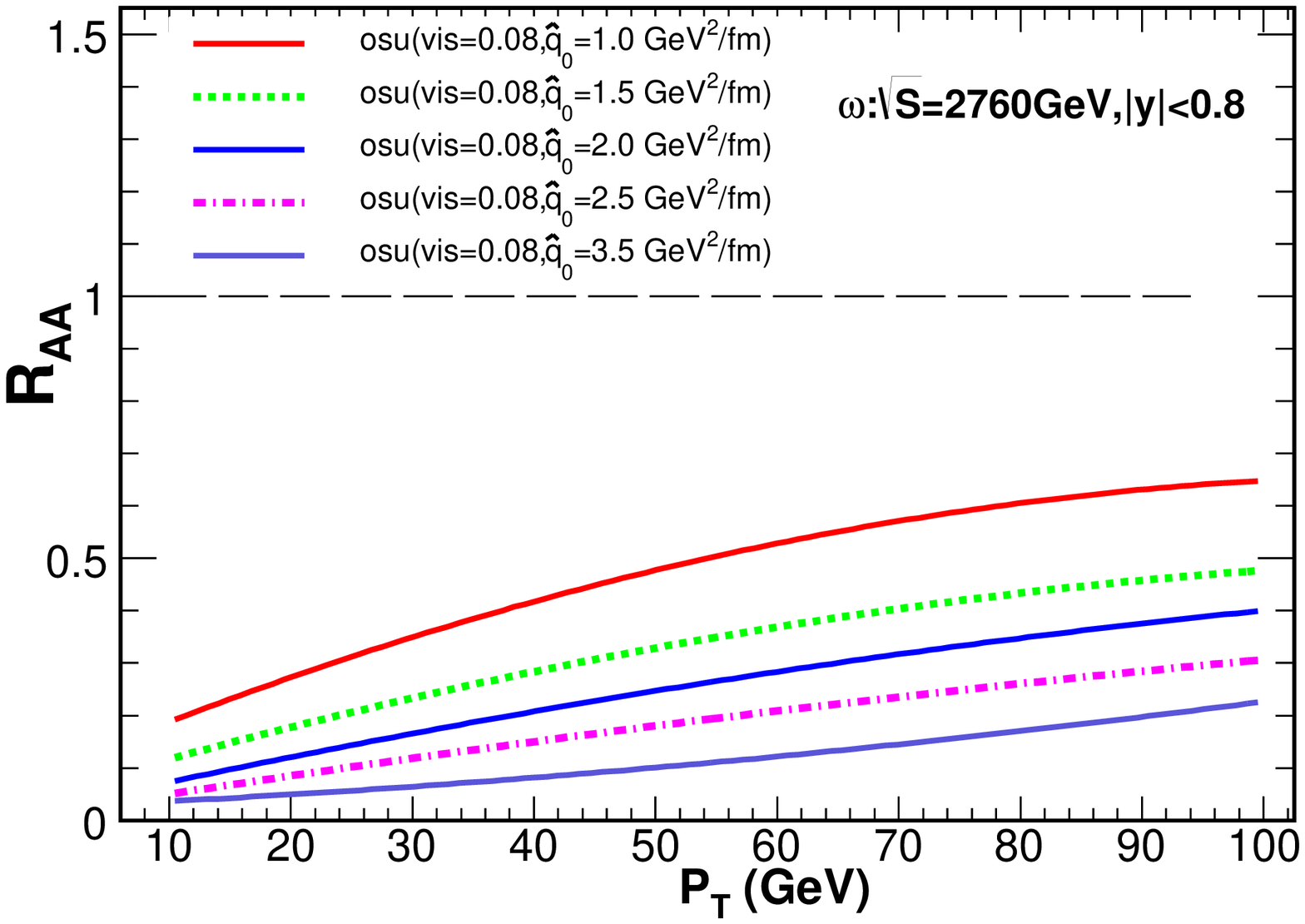}
}
\hspace*{-0.1in}
\vspace*{0.0in}
\caption{Left: Comparison between the PHENIX Data~\cite{Adare:2011ht} of $\omega$ nuclear
modification factor in Au + Au collisions at 200 GeV and numerical
simulations at NLO.
Right: The numerical prediction of $\omega$ nuclear
modification factor in Pb + Pb collisions at 2760 GeV at NLO.}
\label{fig:omgraa}
\end{center}
\end{figure}
\begin{figure}[!t]
\begin{center}
\hspace*{-0.1in}
\vspace*{-0.1in}
\resizebox{0.515\textwidth}{!}{%
\includegraphics{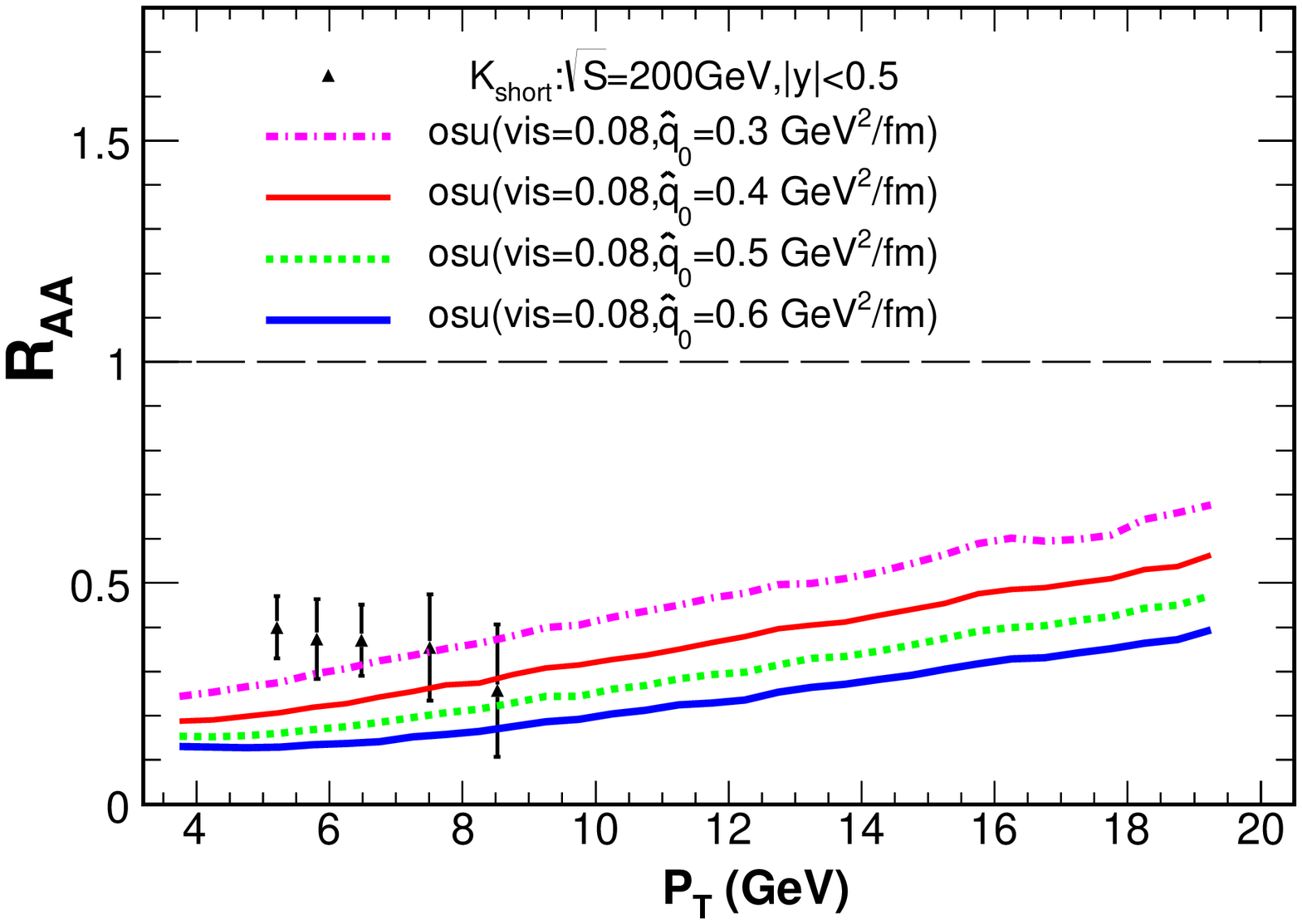}
\includegraphics{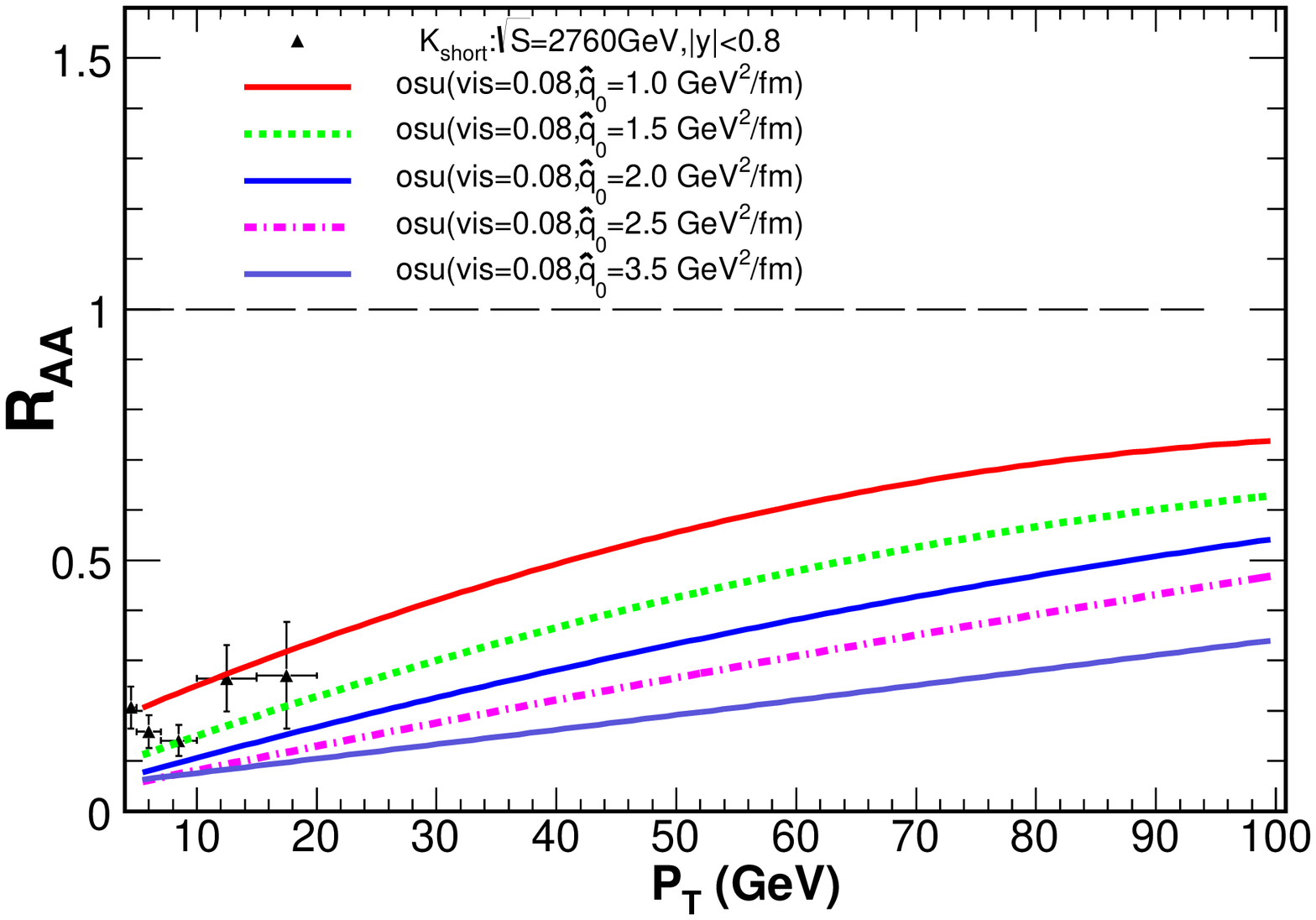}
}
\hspace*{-0.1in}
\vspace*{0.0in}
\caption{Left: Comparison between the STAR Data~\cite{Agakishiev:2011dc} of $K^0_{\rm S}$ nuclear
modification factor in Au + Au collisions at 200 GeV and numerical
simulations at NLO.
Right: Comparison between the ALICE Data~\cite{Adam:2017zbf} of $K^0_{\rm S}$ nuclear
modification factor in Pb + Ob collisions at 2760 GeV and numerical
simulations at NLO.}
\label{fig:ksraa}
\end{center}
\end{figure}

\begin{figure}[!b]
\begin{center}

\resizebox{0.5125\textwidth}{!}{%
\includegraphics{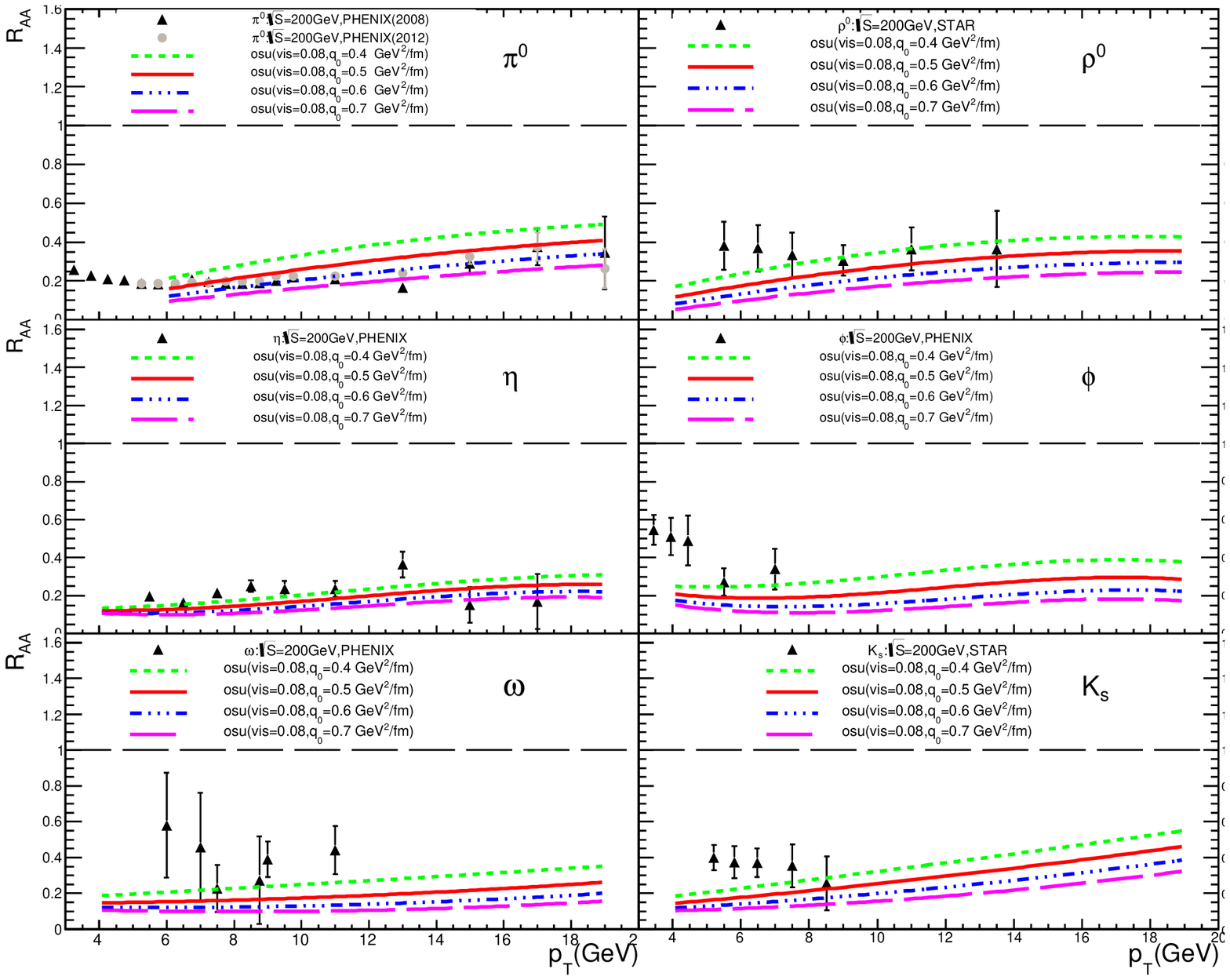}
}

\hspace*{-0.1in}
\caption{Theoretical calculation results of nuclear modification factors $R_{AA}$ as functions of $p_{\rm T}$ at $\hat{q}_0=0.4 - 0.7\rm~GeV^2/fm$
confronted with the RHIC experimental data of $\pi^0$~\cite{Adare:2008qa,Agakishiev:2011dc}, $\rho^0$~\cite{Agakishiev:2011dc},
$K^0_{\rm S}$~\cite{Agakishiev:2011dc}, $\eta$~\cite{Adare:2010dc} and $\phi$~\cite{Adare:2010pt}, $\omega$~\cite{Adare:2011ht}.
}
\label{fig:rhicraa}
\end{center}
\end{figure}
\begin{figure}[!b]
\begin{center}

\resizebox{0.5125\textwidth}{!}{%
\includegraphics{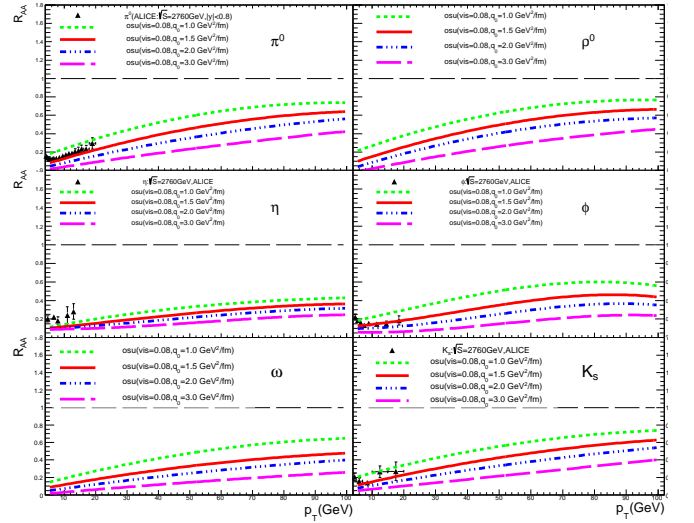}
}

\hspace*{-0.1in}
\caption{Theoretical calculation results of nuclear modification factors $R_{\rm AA}$ as functions of $p_{\rm T}$ at $\hat{q}_0=1.0 - 3.0\rm~GeV^2/fm$
confronted with the LHC experimental data of $\pi^0$~\cite{Adam:2015kca}, $\eta$~\cite{Acharya:2018yhg}, $\phi$~\cite{Adam:2017zbf}, $K^0_{\rm S}$~\cite{Adam:2017zbf}, $\rho^0$ and $\omega$.
}
\label{fig:lhcraa}
\end{center}
\end{figure}

The nuclear modification factor $R_{\rm AA}$ as a function of $p_T$ is calculated as  cross sections in $\rm A+A$ collisions divided by the ones in p+p collision, scaled by the averaged number of binary nucleon-nucleon collisions with a chosen impact parameter $b$~\cite{Adler:2006hu}:
\begin{eqnarray}
R^b_{AB}(p_T, y)=\frac{d\sigma_{AB}^h/dyd^2p_T}{\langle N_{coll}^{AB}(b)\rangle d\sigma_{pp}^h/dyd^2p_T} \ .
\label{eq:raa}
\end{eqnarray}

The theoretical results of $R_{\rm AA}$ at various values of $\hat{q}_0 =0.4 - 0.7 \rm~GeV^2/fm$ for both $\omega$ and $K_{\rm s}$ mesons have been presented in Fig.~\ref{fig:omgraa} and Fig.~\ref{fig:ksraa}.

When obtaining suitable value of $\hat{q}_0$ by comparing the theoretical calculations of $R_{\rm AA}$
for $\omega$ and $K_{\rm s}$ with the corresponding data, there exist two caveats. Firstly, since the data of $\omega$ and $K^0_{\rm S}$ meson at large $p_T$ in A+A collisions are rather limited and with large uncertainty, it is difficult to make a good constrain on $\hat{q}_0$ with these data. Secondly, because we emply the IEBE-VISHNU hydrodynamics model to describe the space-time evolution of the fireball, which gives different information of physics quantities such as temperature and density from those provided by other hydro models such as Hirano hydro description~\cite{Hirano:2001eu,Hirano:2002ds}. Therefore, we could not take advantage of the extracted value of $\hat{q}_0$  in Ref.~\cite{Burke:2013yra,Chen:2010te,Chen:2011vt,Dai:2015dxa,Dai:2017tuy,Dai:2017piq}, where Hirano hydro description has been utilized.

With these two cautions in mind, and realizing that with our model we are now ready to make a systematic study of 6 types of identified mesons such as $\pi^0$, $\eta$, $\rho^0$,  $\phi$,  $\omega$, and $K^0_{\rm S}$ in heavy-ion collisions, it will be of great interest to make a global extraction of jet transport coefficient $\hat{q}_0$ both at RHIC and LHC
by confronting our model calculations (with the IEBE-VISHNU hydro model) against all available data on 6 identified mesons:  $\pi^0$, $\eta$, $\rho^0$,  $\phi$,  $\omega$, and $K^0_{\rm S}$, and then make precise calculations on nuclear modification modification factors of these mesons including  $\omega$, and $K^0_{\rm S}$, as well as yield ratios of these 6 mesons in HIC.

\section{Global Extraction of $\hat{q}_0$ with $R_{\rm AA}$ for Six Identified Mesons}
\label{qhat}
\begin{figure}[!ht]
\begin{center}
\resizebox{0.415\textwidth}{!}{%
\includegraphics{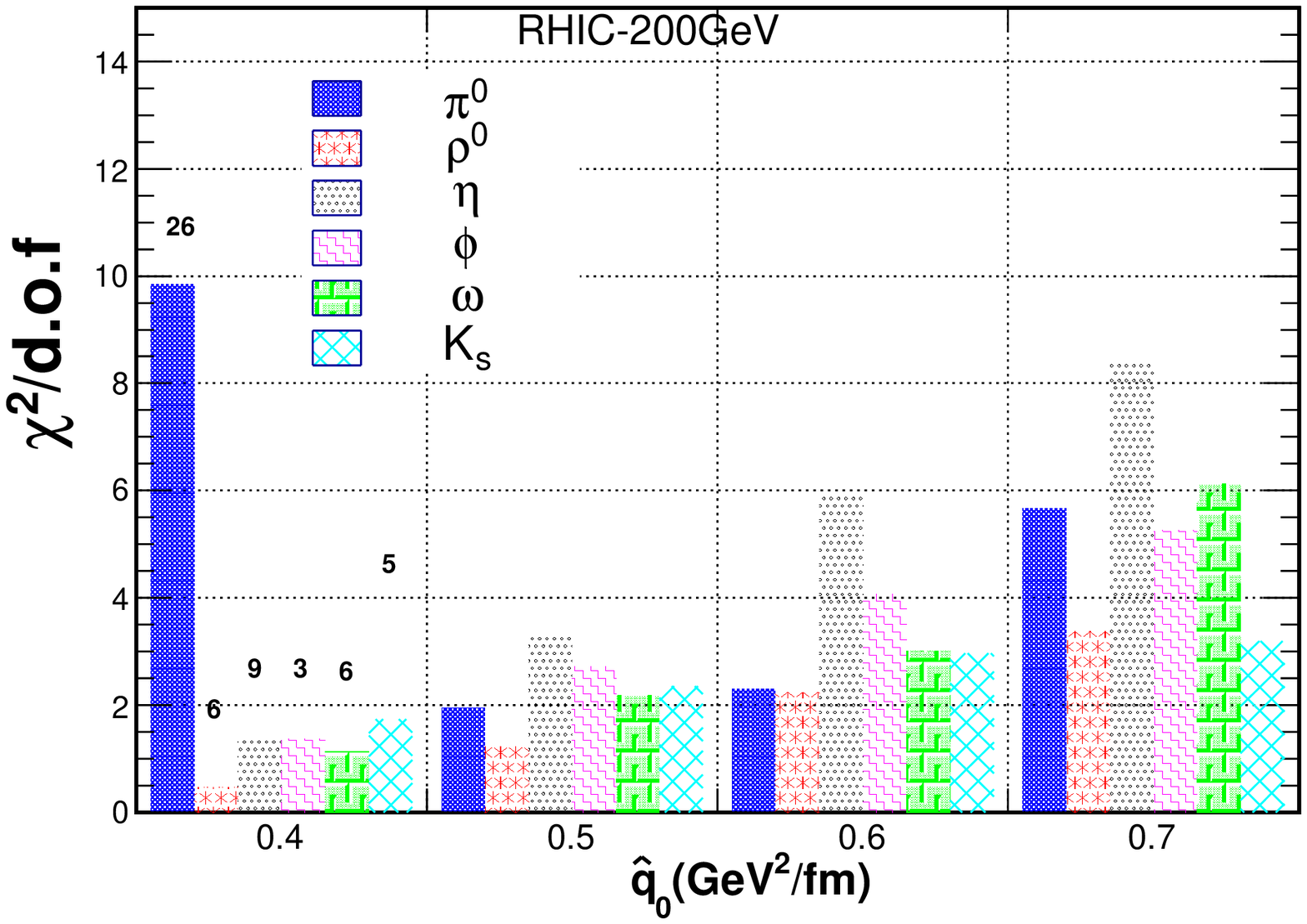}
}
\resizebox{0.415\textwidth}{!}{%
\includegraphics{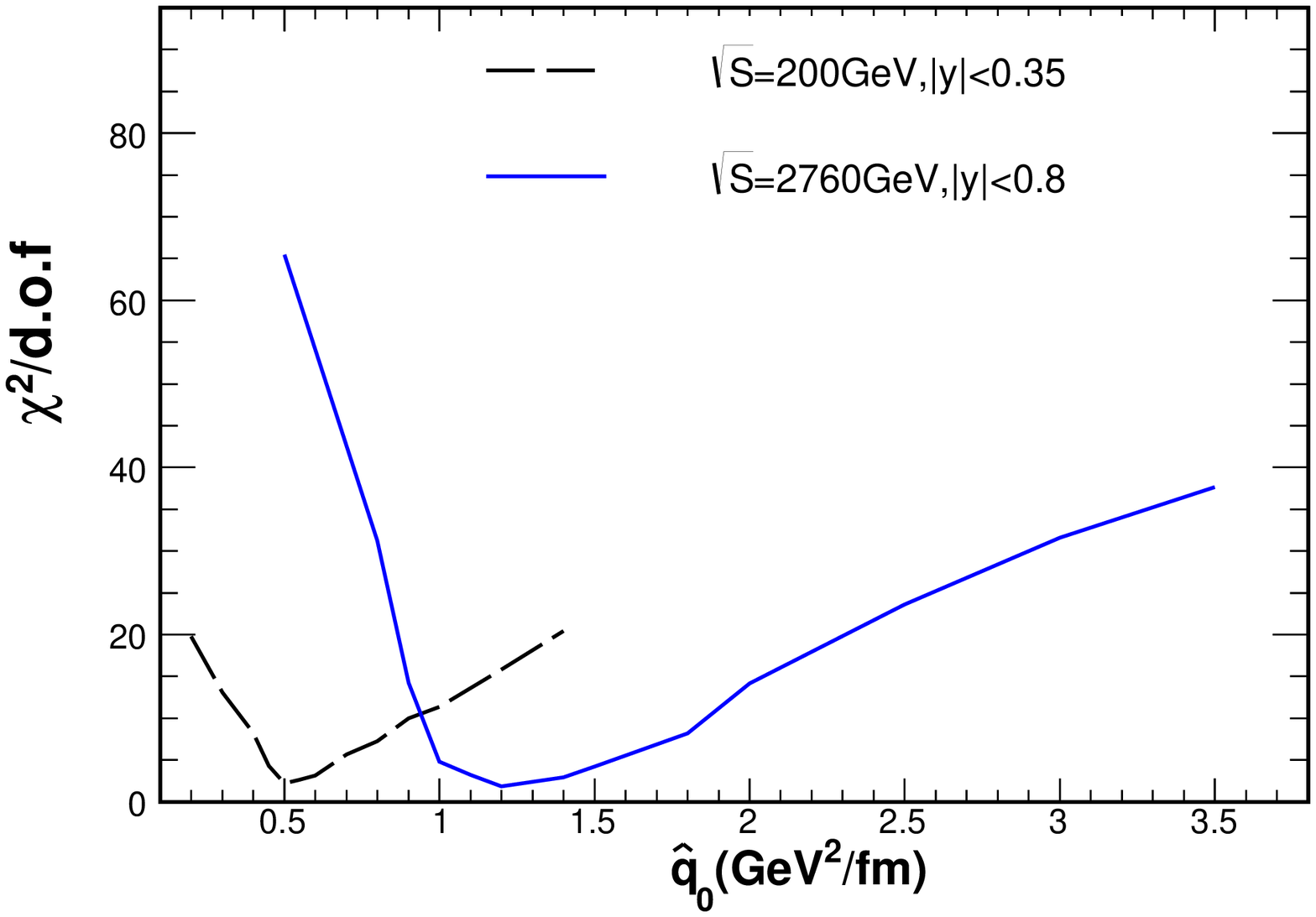}
}
\hspace*{-0.1in}
\caption{Top: demonstration of the $\rm \chi^2/d.o.f$ between the theoretical prediction for $R_{\rm AA}$ of $\pi^0$, $\rho^0$, $\eta$, $\phi$, $\omega$, $K^0_{\rm S}$ at different $\hat{q}_0$ and the current public experimental data at the RHIC $200$~GeV; Bottom: the global $\rm \chi^2/d.o.f$ of $R_{AA}$ taking into account all these 6 identified hadrons at the RHIC $200$~GeV and LHC $2.76$~TeV.
}
\label{fig:chisq}
\end{center}
\end{figure}

We perform a systematic calculation of the $\rm R_{AA}$ of $6$ identified mesons
($\pi^0$, $\rho^0$, $\eta$, $\phi$, $\omega$, $K^0_{\rm S}$) to compare with all their available experimental data both at RHIC and LHC in Fig.~\ref{fig:rhicraa} and Fig.~\ref{fig:lhcraa}.

In order to extract the best value of the initial jet transport parameter $\hat{q}_0$, we perform a $\chi^2$ fit to compare the theoretical results with different $\hat{q}_0$ and the available experimental data of all the different final hadrons.
\begin{eqnarray}
\chi^2({a}) = \sum_i \frac{[D_i-T_i({a})]^2}{\sigma_i^2}
\end{eqnarray}

In the above equations, $D_i$ represents the experimental grids and $T_i$ is our theoretical prediction at input parameter $a$.
$\sigma_i^2$ means the $i-$th systematic and statistical experimental errors. We show in the top panel of Fig.~\ref{fig:chisq}, the derived $\chi^2$ averaged by the number of the compared data points for different final state mesons at various $\hat{q}_0 =0.4 - 0.7\rm~GeV^2/fm$ at RHIC with $\sqrt{s_{\rm NN}}=200$~GeV. In the bottom panel of Fig.~\ref{fig:chisq} we plot the curve $\rm \chi^2/d.o.f$ as a function of $\hat{q}_0$ both at RHIC and LHC, where the minimum of the curve $\rm \chi^2/d.o.f$ with respect to
$\hat{q}_0$ presents the best fit of theory with data. We then observe that at RHIC the minimum point
of $\rm \chi^2/d.o.f$ of all 6 identified mesons gives the best value of $\hat{q}_0=0.5(+0.15/-0.05)\rm~GeV^2/fm$. It is noted that the production of $\pi^0$ will carry the largest weight due to its more abundant data with relatively smaller error bars. We also derive the best value of $\hat{q}_0$ in the Pb+Pb collisions at LHC $\sqrt{s_{\rm NN}}=2.76$~TeV to be $\hat{q}_0=1.2(+0.25/-0.15)\rm~GeV^2/fm$, though the curve $\rm \chi^2/d.o.f$ at LHC is much flatter that at the RHIC. So theoretical results with $\hat{q}_0=1.1-1.4\rm~GeV^2/fm$ should all give decent descriptions on data at LHC.

It is noted that in our current model the extracted values of jet transport coefficient $\hat{q}_0$ both at RHIC and LHC are smaller than that by JET Collaboration in Ref.~\cite{Burke:2013yra} as well as the ones in our preceding calculations~\cite{Dai:2015dxa,Dai:2017tuy,Dai:2017piq}. These differences come mainly from the different hydro models utilized in the studies. We have checked~\cite{Zhang:2018ydt}
that if we employ Hirano hydro description in the calculation, the extracted values of $\hat{q}_0$ from our global fitting will be consistent with those in Ref.~\cite{Burke:2013yra,Dai:2015dxa,Dai:2017tuy,Dai:2017piq}, though the model in this article has the potential to give more precise extraction of jet transport coefficients when more data of identified hadrons in A+A collisions become available in the near future.

\section{Particle Ratios of $\omega/\pi^0$ and $K^0_{s}/\pi^0$ in A+A}
\label{raa}
\begin{figure*}[!t]
\begin{center}

\resizebox{0.415\textwidth}{!}{%
\includegraphics{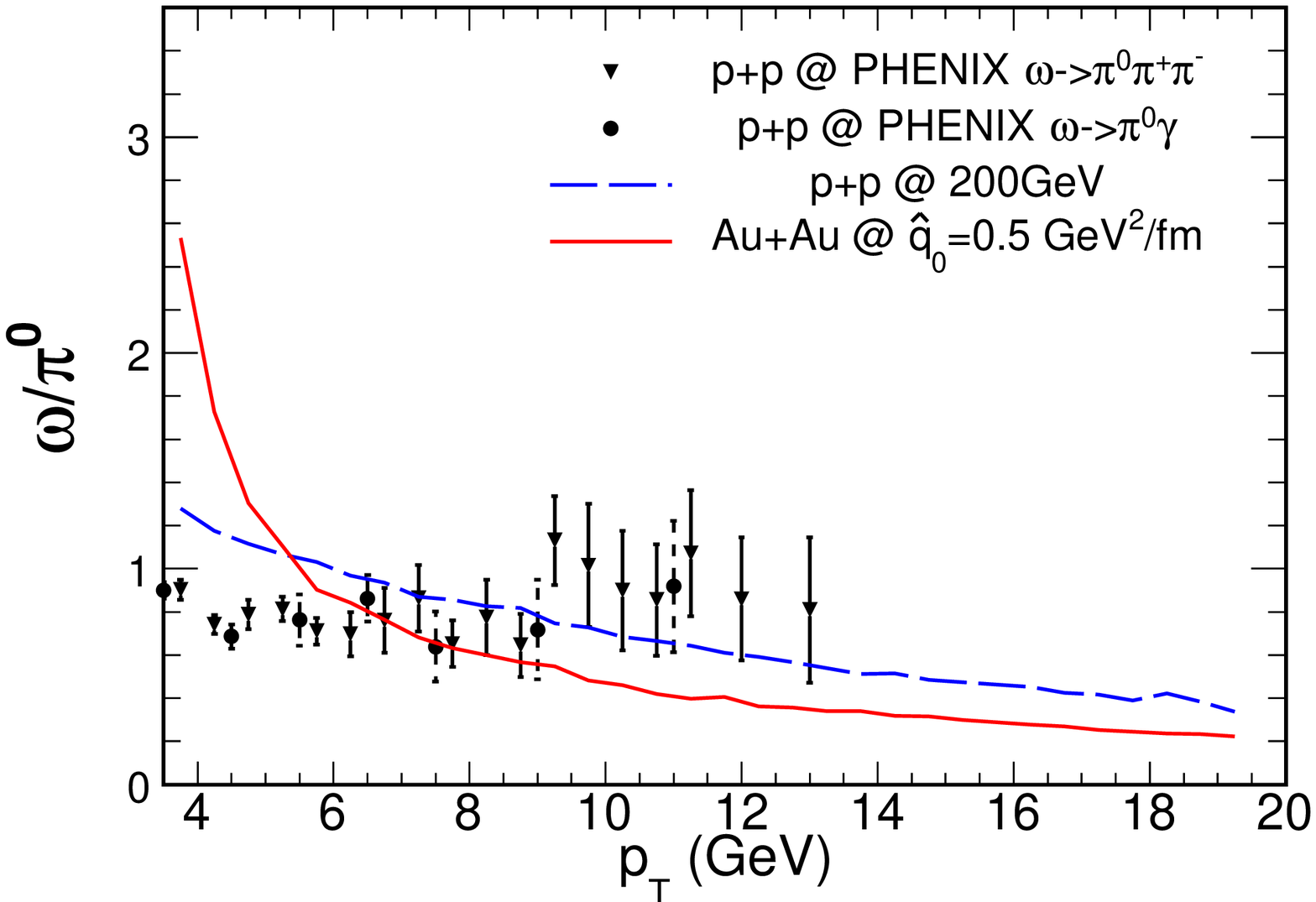}
}
\resizebox{0.415\textwidth}{!}{%
\includegraphics{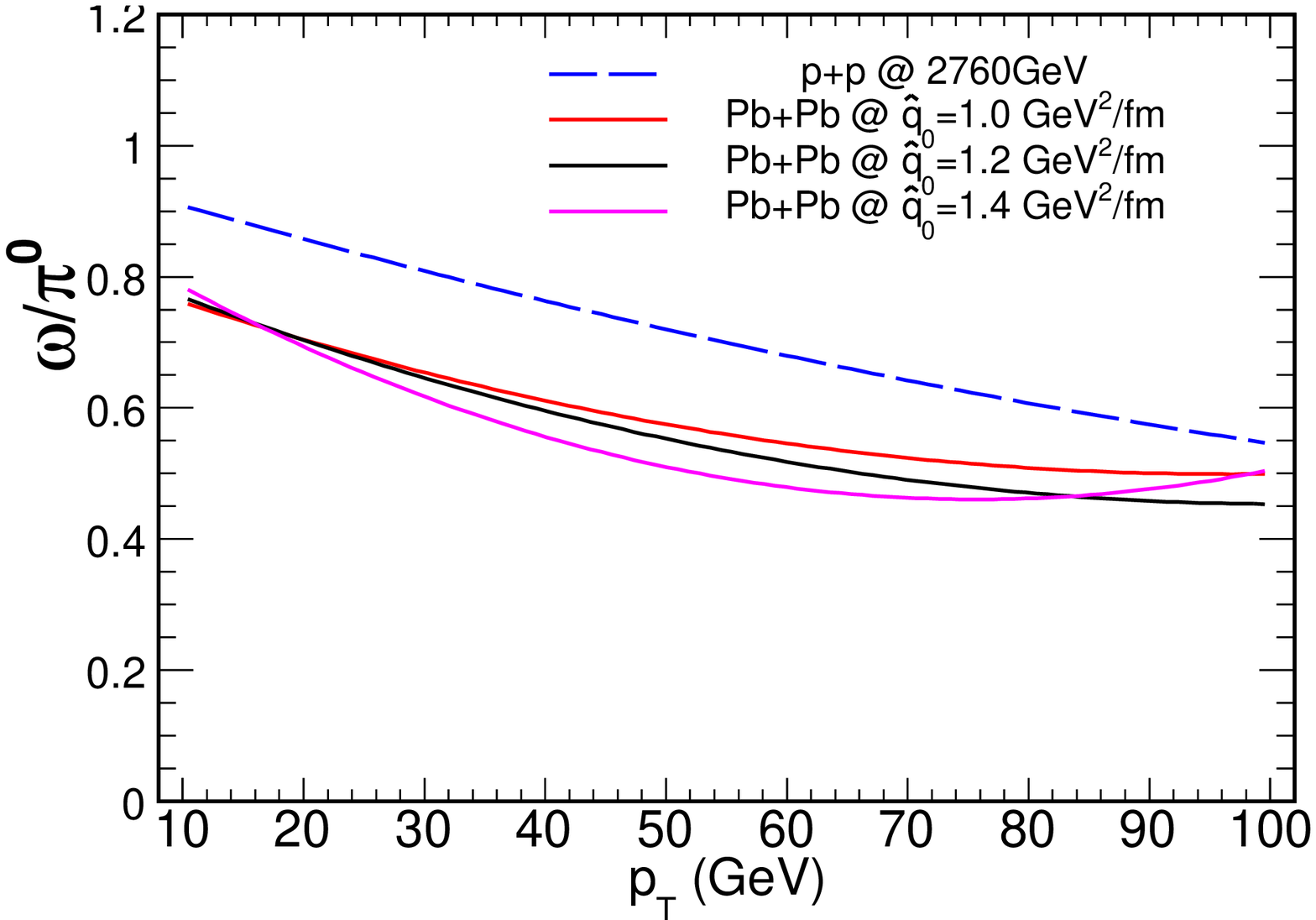}
}

\hspace*{-0.1in}
\caption{Left: the yield ratios of $\omega/\pi^0$ as functions of $p_T$ in p+p and Au+Au collisions with $200$~GeV at RHIC, and PHENIX data in p+p~\cite{Adare:2011ht}; Right: predictions of the yield ratios of $\omega/\pi^0$ as functions of $p_{\rm T}$ in p+p and Pb+Pb collisions with $2.76$~TeV at LHC.
}
\label{fig:omepi}
\end{center}
\end{figure*}
\begin{figure*}[!t]
\begin{center}

\resizebox{0.415\textwidth}{!}{%
\includegraphics{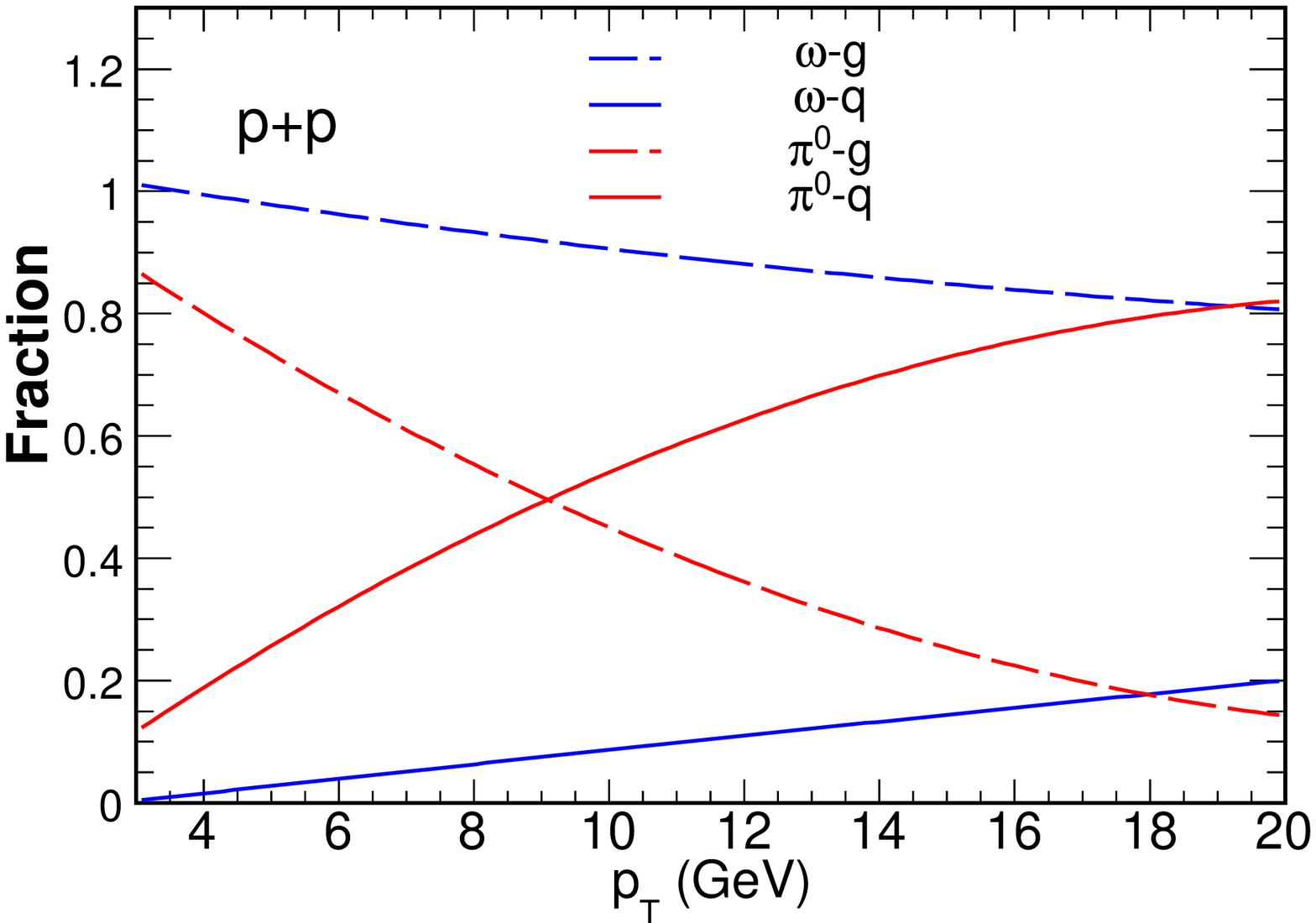}
}
\resizebox{0.415\textwidth}{!}{%
\includegraphics{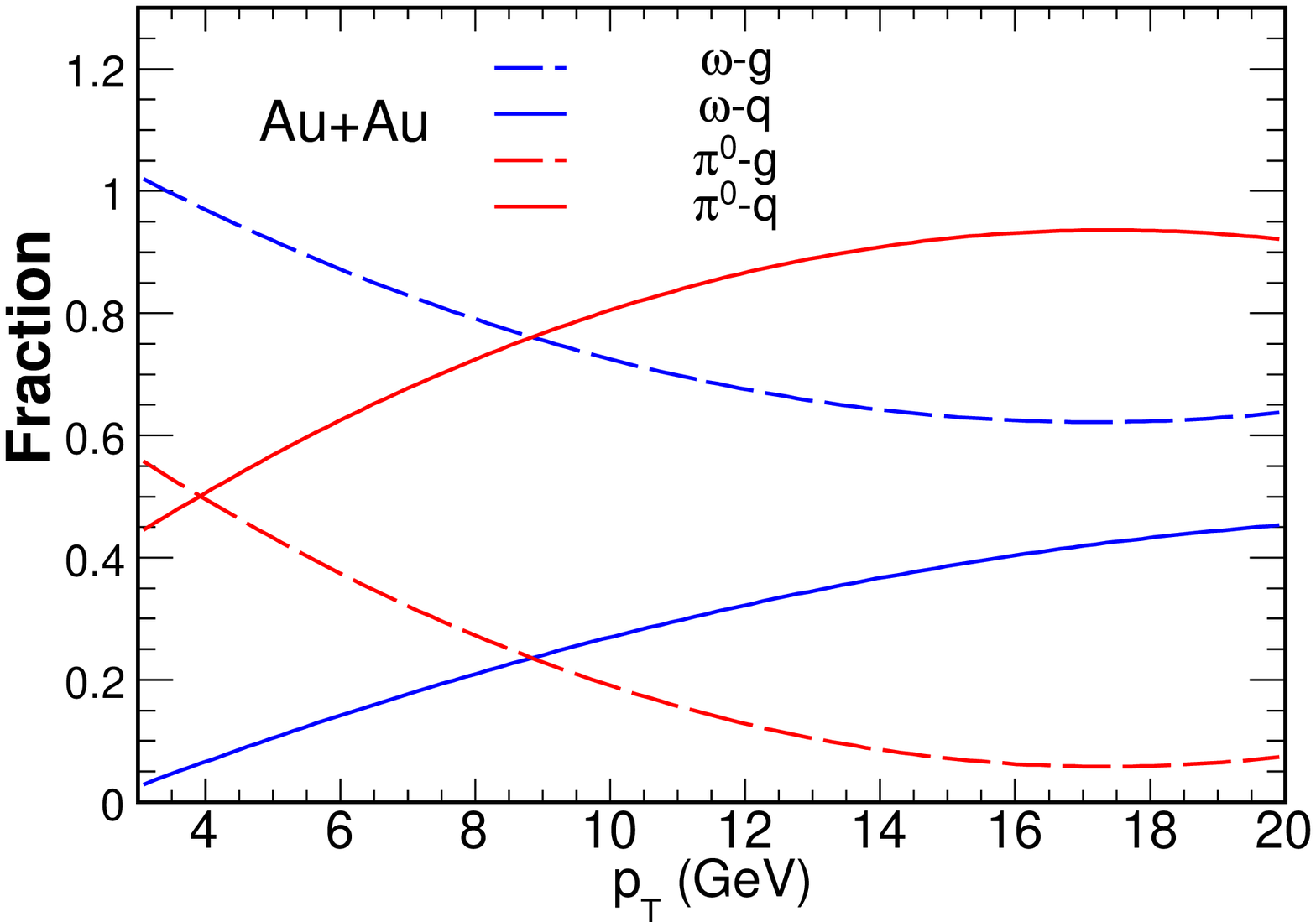}
}

\hspace*{-0.1in}
\caption{Gluon and quark contribution fractions of the total yields of $\omega$ and $\pi^0$
mesons in p+p and Au+Au collisions with $\sqrt{s_{NN}}=200$~GeV at RHIC.
}
\label{fig:omgqg}
\end{center}
\end{figure*}

With the global extracted value of $\hat{q}_0$ discussed in Sec.~\ref{qhat}, we are able to further investigate the particle ratio of $\omega$ and $K^0_{\rm S}$ both at RHIC and LHC.
We first calculate $\omega/\pi^0$ ratio as a function of $p_T$ and show the results in p+p and Au+Au at RHIC in the left panel of Fig.~\ref{fig:omepi}, where the PHENIX experimental data on $\omega/\pi^0$ ratio in p+p are also illustrated.  An enhancement of the ratio in A+A  relative to that in p+p is found in small $p_T$ region, whereas a small suppression in high $p_T$ regime. We also predict the $\omega/\pi^0$ ratio as a function of $p_T$ in p+p and Pb+Pb collisions with $\sqrt{s_{NN}}=2.76$~TeV at LHC.
In Fig.~\ref{fig:omepi} we don't see the overlapping of the curves $\omega/\pi^0$ in p+p and A+A at high $p_T$ regime, as we have observed for the ratios $\eta/\pi^0$~\cite{Dai:2015dxa} and $\rho^0/\pi^0$~\cite{Dai:2017tuy} in p+p and A+A.

To understand this feature deeper, we plot the gluon and quark (fragmentation) contribution fractions to $\omega$ (and $\pi^0$) in p+p and Au+Au at RHIC in Fig.~\ref{fig:omgqg}. One see in the p+p collision at RHIC, the production of $\omega$ is dominated by gluon fragmentation. In the A+A collisions, the gluon suffers more energy loss than the light quark due to its larger color factor,  which decreases the gluon contribution fraction and increase the light quark contribution fraction, but the dominant gluon contribution fraction of $\omega$ $\rm \sim 60\%$ is still observed up to $p_T=20$~GeV in Au+Au at RHIC. On the other hand, $\pi^0$ meson production is light quark fragmentation dominant in p+p, and jet quenching effect further enhance this dominance of light quark fragmentation to $\pi^0$ in A+A. Therefore, we see the yield ratio of $\omega/\pi^0$ in A+A should be suppressed relative to that in p+p due to energy loss effect, and it separates with the one in p+p even at very high $p_T$.
\hspace{-0.2in}
\begin{figure*}[!t]
\begin{center}

\resizebox{0.415\textwidth}{!}{%
\includegraphics{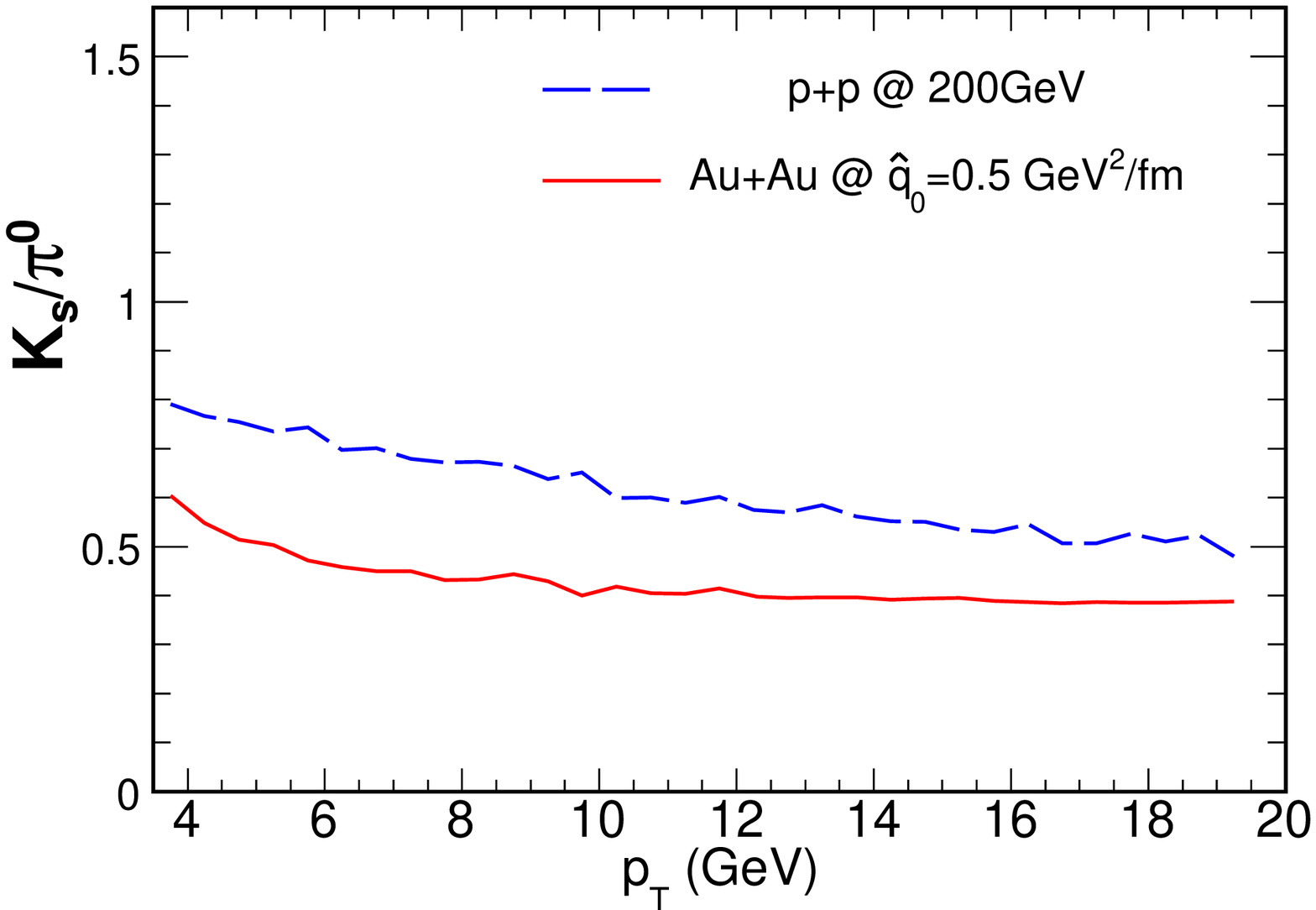}
}
\resizebox{0.415\textwidth}{!}{%
\includegraphics{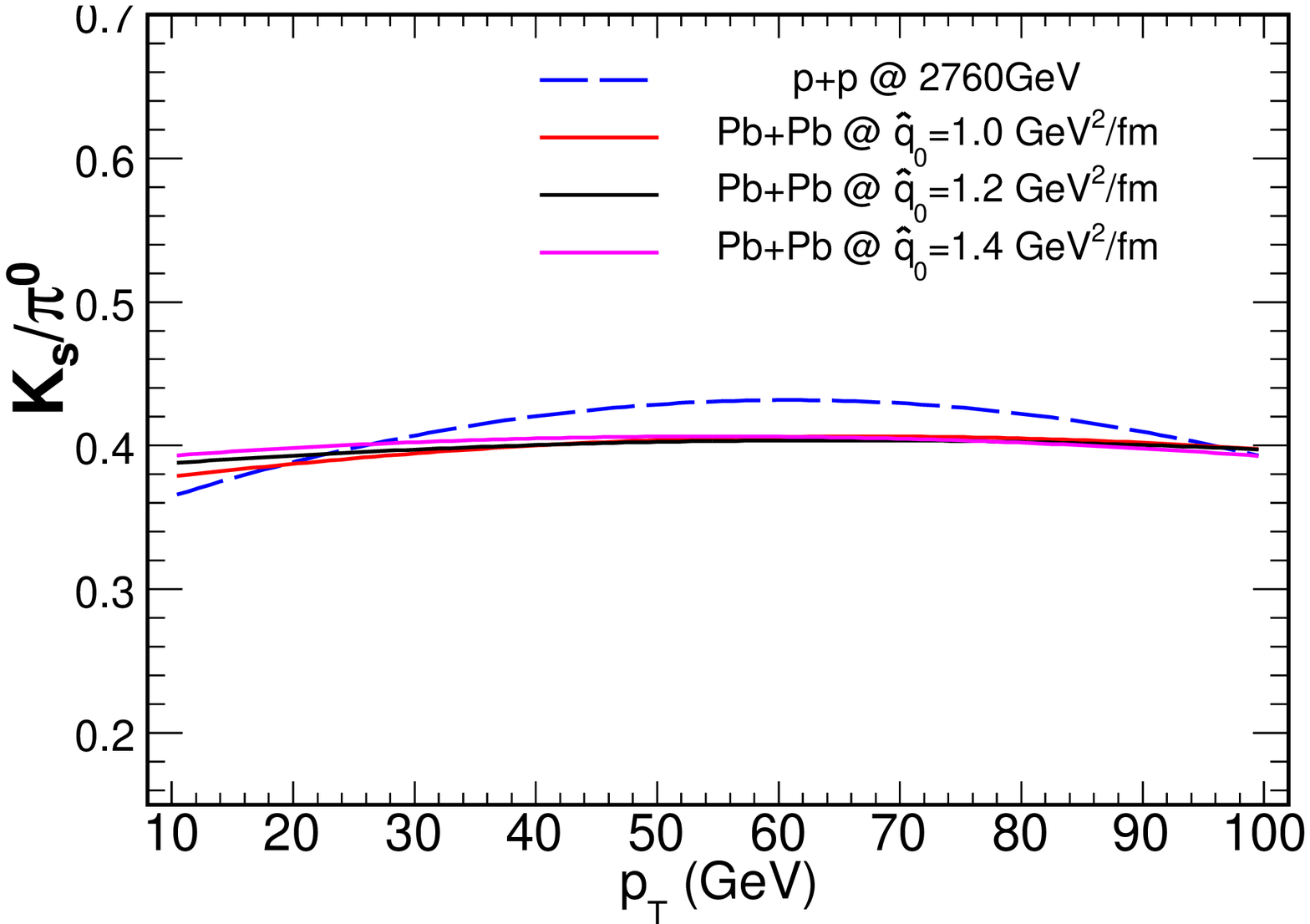}
}

\hspace*{-0.1in}
\caption{Left: the production ratios of $K^0_{\rm S}/\pi^0$ as functions of $p_{\rm T}$ in p+p and Au+Au collisions with $200$~GeV at RHIC; Right: predictions of the production ratios of $K^0_{\rm S}/\pi^0$ as functions of $p_{\rm T}$ in p+p and Pb+Pb collisions with $2.76$~TeV
at LHC.
}
\label{fig:kspi}
\end{center}
\end{figure*}
\hspace{-0.2in}
\begin{figure*}[!t]
\begin{center}

\resizebox{0.415\textwidth}{!}{%
\includegraphics{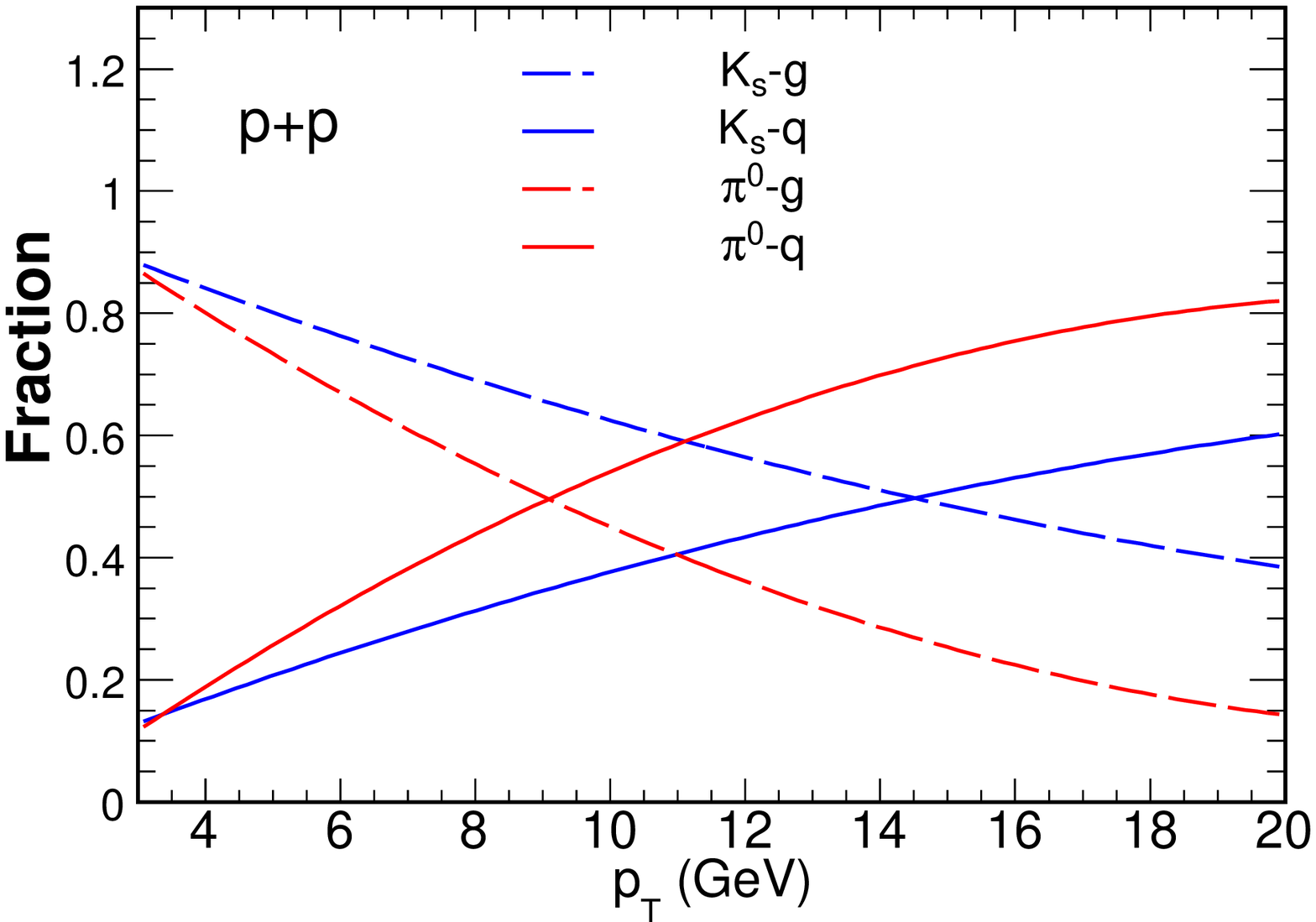}
}
\resizebox{0.415\textwidth}{!}{%
\includegraphics{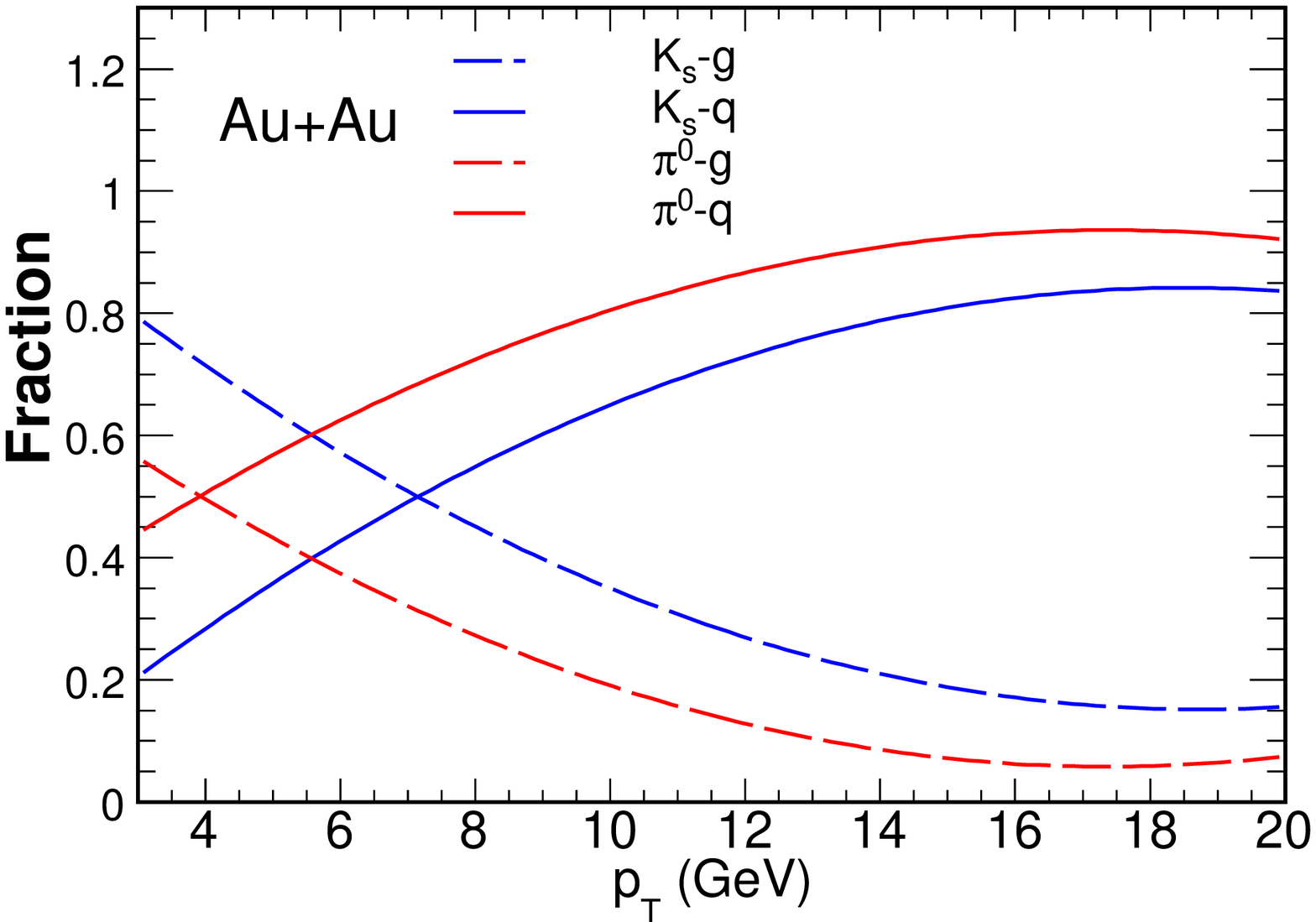}
}

\hspace*{-0.1in}
\caption{Gluon and quark contribution fractions of $K^0_{\rm S}$ and $\pi^0$ yields in $p+p$ and $Au+Au$ collisions with $\sqrt{s_{NN}}=200$~GeV at RHIC.
}
\label{fig:ksqg}
\end{center}
\end{figure*}

We also compute the $K^0_{\rm S}/\pi^0$ ratio as a function of $p_T$ both at RHIC and LHC in Fig.~\ref{fig:kspi}. We find the curves in A+A and in p+p are approaching to each other with $p_T$ increasing,  and an obvious coincidence of these two curves is seen at LHC. We show the gluon and quark contribution fractions to $K^0_{\rm S}$ ( and $\pi^0$ meson) as functions of $p_T$
in Fig.~\ref{fig:ksqg}. We find in p+p collision, the productions of both $K^0_{\rm S}$ and $\pi^0$ at very large transverse momenta are dominated by quark fragmentation. In A+A collisions, the gluon contribution shall be further suppressed because gluon generally loses more energy. Thus, both in p+p and A+A collisions, the ratio $K^0_{\rm S}/\pi^0$ should be largely determined by the ratio of quark FFs for $K^0_{\rm S}$
$(D^{K^0_{\rm S}}_q(z_h, Q^2))$ to quark FFs for $\pi^0$
$(D^{\pi^0}_q(z_h,Q^2))$ at very high $p_T$, where these FFs vary slowly with the momentum fraction $z_h$, very similar to the case of $\eta_0/\pi^0$ at high $p_T$~\cite{Dai:2015dxa}. Even though in A+A collisions, jet quenching effect can shift $z_h$ of quark FFs, if quark FFs have a rather weak dependence on $z_h$ and $p_T$, we can see at very high-$p_T$ regime the curves for $K^0_{\rm S}/\pi^0$ in A+A and p+p are coming close to each other, and even coincide at LHC.

\section{Summary}
\label{summary}
In summary, we obtain the NLO FFs of $\omega$ meson in vacuum by evolving the rescaled $\omega$ FFs from a broken SU(3) model at a starting scale $\rm Q^2_0=1.5~GeV^2$, and directly employ NLO $K^0_{\rm S}$ FFs in vacuum from the AKK08 parameterizations, the numerical simulation of productions of both $\omega$ and $K^0_{\rm S}$ matches well with the experimental data in p+p reactions. With the IEBE-VISHNU hydro profile of the QCD medium, we calculate the nuclear modification factors of $\omega$ and $K^0_{\rm S}$ meson as well as $\pi^0$, $\eta$, $\phi$, $\rho^0$ in A+A collisions both at RHIC and LHC, including jet quenching effect in higher-twist approach. The global extraction of jet transport parameter $\hat{q}_0$ are made, with comparison of the theoretical calculation and the experimental data of all six identified mesons: $\pi^0$, $\rho^0$, $\eta$, $\phi$, $\omega$, $K^0_{\rm S}$ . Furthermore, we predict the yield ratios of $\omega/\pi^0$ both at RHIC and LHC, and a fairly good agreement of the theoretical results and experimental data is found at RHIC. Theoretical predictions of $K^0_{\rm S}/\pi^0$ ratios as functions of $p_T$ at RHIC and LHC are presented as well.

\vspace*{.3cm}
{\bf Acknowledgments:}  The research is supported by the NSFC of China with Project Nos. 11435004 and 11805167, and partly supported by the Fundamental Research Funds for the Central Universities, China University of Geosciences (Wuhan) (No. 162301182691).
%

\newpage
\appendix
\section{:~Input parameters for $\omega$ in broken SU(3) model}
\label{inputs}
\begin{table}[htbp]
\centering
\begin{tabular}{|c|c|c|} \hline
~~~~~~~~~~~ &  ~~~~~~~~~~    & ~~~~~~~~~$\omega$~~~~~~~~~  \\ \hline
$V$     &$a$ & 2.16      \\ \hline
        &$b$ & 0.52       \\ \hline
        &$c$ & 1.24       \\ \hline
        &$d$ & 0.27       \\ \hline
        &$e$ & -0.16     \\ \hline
$\gamma$    &$a$ & 2.97      \\ \hline
        &$b$ & -0.48     \\ \hline
        &$c$ & 5.48       \\ \hline
        &$d$ & -0.09      \\ \hline
        &$e$ & 1.25       \\ \hline
$D_g$   &$a$ & 11.67      \\ \hline
        &$b$ & 0.745     \\ \hline
        &$c$ & 3.14       \\ \hline
        &$d$ & -0.13     \\ \hline
        &$e$ & -0.21     \\ \hline
$\lambda$     &       & 0.07     \\ \hline
$\theta $     &     & 40.49    \\ \hline
$f_g$   &   & 1.00     \\ \hline
$f_{sea}$ & & 0.99     \\ \hline
$f_1^u$   &   & 0.05    \\ \hline
$f_1^s$   &   & 0     \\ \hline
\end{tabular}
\caption{The parameters for the input fragmentation functions of $\omega$ at the initial scale $Q^2 = 1.5 \ GeV^2$
based on broken SU(3) model in our calculation.}
\label{tab:inputs}
\end{table}

\end{document}